\documentclass[notitlepage,longbibliography]{revtex4-2}

\usepackage{amsmath,amssymb}
\usepackage{latexsym,bbm}
\usepackage{mathdots}
\usepackage{bm}

\usepackage[bookmarks=false,colorlinks=true]{hyperref}
\usepackage{orcidlink}

\newcommand{\sfrac}[2]{{\textstyle\frac{#1}{#2}}}

\newcommand{\Hgencoul}{{\mathcal{H}}^\text{gen}_\gamma}

\newcommand{\gen}{\mathcal{L}}
\newcommand{\cf}{\mathcal{S}}

\newcommand{\deform}[1]{\mathcal{H}_g#1}

\begin{document}

\title{Dynamical symmetries of the Calogero--Coulomb model}

\author{Tigran Hakobyan\,\orcidlink{0000-0003-2004-9920}}
\email{tigran.hakobyan@ysu.am}
\email{hakob@yerphi.am}
\affiliation{Yerevan State University, 0025 Yerevan, Armenia}
\affiliation{A.~Alikhanyan National Science Laboratory (YerPhI), 0036 Yerevan, Armenia}

\date{\today}

\begin{abstract}
We construct the dynamical symmetry of the $N$-particle quantum Calogero model with particle exchange in
a confining Coulomb field.
This symmetry is governed by the  algebra $so(N+1,2)$, deformed by exchange (Dunkl) operators,
with its invariant sector generated by the Dunkl angular momentum tensor and
the modified Laplace--Runge--Lenz vector.
The equidistant analogue of the Hamiltonian, with a linear spectrum,
is expressed in terms of the conformal
subalgebra $so(1,2)$.
In addition, the wave functions of the Calogero--Coulomb Hamiltonian are classified
into infinite-dimensional lowest-weight
$so(1,2)$ multiplets.
\end{abstract}

\maketitle

\section{Introduction}
Exactly solvable systems play a fundamental role in quantum physics, providing explicit realizations of hidden symmetries and serving as benchmarks for nonperturbative methods.
Among these systems, the Calogero model, which describes one-dimensional identical particles with inverse-square interactions, is a prominent example \cite{calogero-1,calogero-2,moser}.
More general inverse-square potentials based on finite reflection groups,
as well as hyperbolic, trigonometric, and elliptic functions, spin-exchange interactions,
and related extensions, are also integrable (see the reviews \cite{rev-olsh-2,rev-poly,book-etin,book-arut}
and references therein).
Moreover, the rational Calogero models possess a complete set of constants of motion, rendering them maximally superintegrable \cite{woj83,kuznetsov,gonera}. The   (super)integrability is robust under the inclusion of certain external potentials
 and persists for the angular parts of the associated Hamiltonians, as well as for systems defined on surfaces of constant curvature \cite{hak14-1,runge,flp,fh15,fh22}.

On the other hand, the inverse-square interaction potential complicates the structure of Calogero models
but does not affect the  (super)integrability of the system \cite{hak14-1}.
For instance, the full symmetry of the Calogero model --- originally formulated with an oscillator
confining potential --- can be regarded as a $g$ deformation of the $u(N)$ unitary algebra,
which encodes the invariants of the $N$-dimensional isotropic oscillator \cite{turbiner,fh15};
see also Refs.~\cite{feigin23,vrabec25} for a quantum-group extension.
Here, the parameter $g$ specifies the coupling constant of the inverse‑square interaction.
This correspondence extends to the dynamical symmetry, which encompasses
also ladder operators that raise or lower states along the energy scale \cite{hak23}.

It is well known that the $N$-dimensional  confining Coulomb model is also superintegrable
characterized by the symmetry group  $SO(N+1)$,  generated by the angular momentum and the
Laplace--Runge--Lenz vector.  The   spectrum of the quantum system, however, is not equidistant.
At first glance,  this appears  to preclude the existence of spectrum-generating
operators. Nevertheless, a modified system can  be constructed that shares the same stationary
states while exhibiting a linear spectrum.
The equidistant Hamiltonian possesses a dynamical symmetry described by the conformal group
$SO(N+1,2)$ in $(N+1)$-dimensional Minkowski space.
This structure was first demonstrated in the case of the hydrogen atom \cite{barut67,barut71}.

The energy spectrum of the Coulomb model with a rational Calogero potential,
is known to possess a huge degeneracy  \cite{khare}.
At the same time, the Calogero--Coulomb system,
including its extensions to constant-curvature spaces, is superintegrable
\cite{hak14-1}.
The corresponding constants of motion, including an analogue of
the Laplace--Runge--Lenz vector, have been constructed \cite{runge,hak16-1,hak16-2}.
Moreover, the complete set of conserved quantities of the (confining) quantum Calogero--Coulomb
problem forms a $g$-deformation of the $so(N + 1)$ algebra \cite{fh22}.

In this article, we construct and analyze the dynamical symmetries of the quantum Calogero--Coulomb system.
Following many of the earlier developments in Calogero-type models, our approach employs the
exchange (Dunkl) operator formalism, which incorporates particle-exchange effects with coupling constant $g$
into covariant differential \cite{dunkl}. This formulation provides a powerful and algebraically transparent framework for systematically deriving solutions, integrals of motion, symmetry algebras, and ladder operators.

We demonstrate that the dynamical symmetry (spectrum-generating symmetry) of the
quantum Calogero--Coulomb system is described by an exchange-operator deformation of the conformal algebra
$so(N+1,2)$. Within this structure, a three-dimensional conformal subalgebra $so(1,2)$ emerges,
which contains the modified Calogero--Coulomb Hamiltonian with  linear spectrum.
For the modified system, we construct a deformed Laplace--Runge--Lenz vector, distinct
from the analogous vector obtained for the original model \cite{runge}. Together with the
Dunkl angular momentum, this vector generates a complete symmetry algebra of the equidistant
Calogero--Coulomb model.
This algebra represents an exchange-operator deformation of the $so(N+1)$ algebra,
with its Casimir element expressed in terms of the Hamiltonian, and it retains a structure closely related to that of the original system \cite{fh22}.

The commutation relations among the dynamical symmetry generators are derived.
They incorporate particle-exchange operators directly into the structure constants
and are expressed compactly in matrix form.
Furthermore, we establish the equivalence between the quadratic Casimir elements of
the $so(1,2)$ and Dunkl angular momentum algebras,
and we calculate the numerical value of the Casimir element of the dynamical symmetry algebra.

Finally, we construct the wave functions of the Calogero--Coulomb Hamiltonian,
both with and without particle exchange. These wave functions are classified
according to the infinite-dimensional lowest-weight irreducible representation of the
$so(1,2)$ conformal algebra.
The conformal spin of the multiplet is parameterized by the orbital quantum number,
while the states are labeled by the radial quantum number.

The paper is organized as follows.
In Sect.~\ref{sec:model}, we briefly review the Calogero--Coulomb model and
its symmetries within the Dunkl-operator formalism.
In Sect.~\ref{sec:conf}, we construct and analyze  the $so(1,2)$ conformal subalgebra
together with an equidistant analogue of the Calogero--Coulomb
Hamiltonian.
In Sect.~\ref{sec:dyn}, we develop  a Dunkl-deformed analogue  of the full conformal algebra
$so(N+1,2)$, whose rotated realization describes the  dynamical symmetry
of the Calogero--Coulomb Hamiltonian.
In Sect.~\ref{sec:conf-struc}, using deformed spherical harmonics,
we construct the eigenfunctions of the Calogero--Coulomb system
 and study their $so(1,2)$ multiplet  structure.
The results are summarized and discussed in Sect.~\ref{sec:concl}.
\ref{app:comm} presents the derivation of the commutation relations
among the generators of the dynamical symmetry algebra, while \ref{app:Asq}
contains explicit calculations of the squares of three vectors,
including the Laplace--Runge--Lenz vector, that enter the algebra.

\section{Calogero--Coulomb system}
\label{sec:model}
In this section, we briefly review the exchange-operator formalism and its application to the
symmetries of the Calogero--Coulomb model with particle exchange.

\subsection{Hamiltonian via exchange (Dunkl) operators}
The Hamiltonian of the $N$-particle
quantum Calogero model in external Coulomb potential
\cite{khare} has the following form:
\begin{equation}
\label{coul}
\tilde H_\gamma  =\sum_{i=1}^N\frac{p_i^2}{2m}  +
 \sum_{i<j} \frac{g(g\mp \hbar)}{m(x_i-x_j)^2}-   \frac{\gamma}{r},
 \qquad
 r =\sqrt{x^2},
\end{equation}
where $p_i=-\imath \hbar \partial_i$ is the momentum operator, $x^2=\sum_{i=1}^N x_i^2$, and the $-/+$ sign in the interaction potential corresponds
to identical bosons/fermions, respectively. For $\gamma=0$,
the above Hamiltonian reduces to the unconfined Calogero model \cite{calogero-1,calogero-2},
also known as the Calogero--Moser system \cite{moser}.
Throughout this article, we consider positive coupling constants, $g,\gamma>0$, and set
 the mass and Plank's constant to unity, $m=\hbar=1$.

An elegant approach to quantum Calogero-type models is based on the Dunkl operators,
whose components mutually commute \cite{dunkl};
see Refs.~\cite{Heckman1997,rev-rosler,rev-chalykh,book-dunkl} for reviews:
\begin{align}
\label{du}
&\nabla_i=\partial_i-\sum_{j\ne i} \frac{g}{x_i-x_j}s_{ij},
\\
\label{Sij}
&[\nabla_i,\nabla_j]=0,
\qquad
[\nabla_i,x_j]=S_{ij}:=\begin{cases}
-g s_{ij} & i\ne j,
\\
1+g\sum_{k\ne i} s_{ik} & i=j.
\end{cases}
\end{align}
In the above equations, the operator $s_{ij}$ permutes the $i$-th and $j$-th
particles. The Dunkl operator can be viewed as a covariant
derivative with a flat connection and a nonlocal imaginary field with particle
exchanges. Its commutation relations with the coordinates also involve exchange
operators and reduce to the canonical commutation relations in the free-coupling limit,
$\lim_{g\to 0} S_{ij}=\delta_{ij}$.

The algebra \eqref{Sij} is often referred to in the literature as the
$S_N$-extended Heisenberg  algebra \cite{brink'}.
The corresponding abstract algebraic structure is known as
the rational Cherednik algebra associated with the symmetric group \cite{cherednik,etingof,etingof-ma}.
The Dunkl operator together with the coordinate generate a  particular representation
of this algebra.

Pursuing further the analogy with particle motion in an external field, one can define
a covariant momentum operator $\pi_i$. The corresponding Hamiltonian includes a Calogero-type potential
that, however, involves particle exchanges \cite{poly92,brink}.
In the presence of a Coulomb potential, this leads to the Calogero--Coulomb Hamiltonian with particle exchange
\cite{hak14-1,runge}:
\begin{equation}
\label{Hgamma}
H_\gamma=\sum_{i=1}^N\frac{\pi_i^2}{2}-\frac{\gamma}{r}=\sum_{i=1}^N\frac{p_i^2}{2}
+ \sum_{i<j}\frac{g(g-s_{ij})}{(x_i-x_j)^2}-\frac{\gamma}{r},
\qquad
\pi_i=-\imath \nabla_i.
\end{equation}
In what follows, we adopt the shorthand notations $\pi^2=\sum_i \pi_i^2$ and $p^2=\sum_i p_i^2$,
denoting the squared covariant momentum and the squared canonical momentum, respectively.
 For  identical bosons ($s_{ij}=1$) or fermions ($s_{ij}=-1$),  the
 Hamiltonian $H_\gamma$ reduces to the convensional Calogero--Coulomb model \eqref{coul}.
 Calogero-type Hamiltonians that incorporate particle exchange within the inverse-square interaction
are commonly referred to in the literature as generalized Calogero models.
In this article, we focus primarily on such systems and therefore designate
the particle-exchange Hamiltonian \eqref{Hgamma} simply as the Calogero--Coulomb Hamiltonian.

Let us briefly outline Dunkl operators and their applications in physics.
The exchange-operator formalism is widely used to establish the quantum integrability of many-body systems
with inverse-square interactions and to construct stationary wave functions algebraically \cite{poly92,brink}.
Note that the Calogero model   is also integrable via the quantum inverse scattering method
\cite{hikami}, and a direct correspondence between the two approaches has been identified \cite{chalykh19}.
A trigonometric extension of exchange operators leads to the Calogero--Sutherland model \cite{hechman91,cherednik}, while Dunkl and Cherednik operators admit elliptic generalizations \cite{veselov94,cherednik95}, which establish the integrability of the elliptic Calogero--Moser model \cite{cherednik95}.
For integrable spin chains with long-range interactions, Dunkl operators frozen at equilibrium yield commuting invariants of motion \cite{poly93,minahan93,chalykh24}.
These invariants also arise in the semiclassical limit \cite{talstra95,mathieu01,hak-chain}. More recently, a general framework for quantum integrable systems on integrable classical backgrounds --- emerging naturally in the semiclassical limit --- has been developed \cite{resh25}.
Calogero models with particle and spin exchange exhibit hallmark features of integrability, including Yangian symmetry, a Lax pair formulation, and the presence of both an $R$-matrix and a monodromy matrix satisfying the Yang--Baxter equation \cite{bernard93,serban24,maria25}.
Furthermore, for integer values of the coupling constant $g$, the rational model possesses
nonlinear supersymmetry, as established within the exchange-operator formalism \cite{mikhail}.

\subsection{Symmetries of Calogero--Coulomb model}

 Using appropriately  the covariant momentum \eqref{Hgamma},
 one can construct Dunkl-operator deformations
 of other physical quantities. In particular, the Dunkl angular momentum
 tensor  \cite{feigin,kuznetsov,fh15}
 and Laplace--Runge--Lenz vector \cite{runge}  are given, respectively, by the following
 expressions:
\begin{align}
\label{Lij}
L_{ij}&=x_i\pi_j-x_j\pi_i,
\\
\label{Agamma}
 A^\gamma_i&=\frac1 2 \sum_{j=1}^N\left\{ L_{ij},\pi_j\right\}
-\frac\imath2[\pi_i,S] -\frac{\gamma x_i}{r}.
\end{align}
Here, the upper index of $A^\gamma_i$, as the lower index of $H_\gamma$,
indicates the dependence on the Coulomb coupling constant.
The curly brackets denote  the anticommutator, $\{a,b\}:=ab+ba$.
The (rescaled) invariant element of the permutation-group algebra appearing in Eq.~\eqref{Agamma}
is defined as follows:
\begin{equation}
 \label{S}
 S=\sum_{i<j} S_{ij}=-g\sum_{i<j} s_{ij},
 \qquad
 [S,s_{ij}]=0.
\end{equation}
The Hamiltonian preserves both of the quantities defined above:
\begin{equation}
\label{comHLA}
[H_\gamma,L_{ij}]=[H_\gamma,A^\gamma_i]=0.
\end{equation}
In the $g=0$ limit,  both systems \eqref{coul} and \eqref{Hgamma}
reduce to the $N$-dimensional Coulomb  model. Accordingly, the operators
\eqref{Lij} and \eqref{Agamma} reduce to the standard angular momentum
and Laplace--Runge--Lenz components.

The commutation relations among the components $L_{ij}$ and $A^\gamma_k$
are  \cite{kuznetsov,fh15,runge}:
\begin{align}
\label{comLL}
[L_{ij},L_{kl}]&=
\imath(L_{il}S_{kj} +  L_{jk}S_{li} - L_{ik}S_{lj} -  L_{jl}S_{ki}),
\\[1mm]
\label{comLA}
[L_{ij}, A^\gamma_k]&=   \imath(A^\gamma_i S_{jk} - A^\gamma_j S_{ik}),
\\[1mm]
\label{comAA}
[A^\gamma_i,A^\gamma_j]&= -2\imath H_\gamma L_{ij}.
\end{align}
These are $g$-deformations of the commutations between
the standard  angular momentum  and Laplace--Runge--Lenz invariants.
However, unlike the $g=0$ case, the above relations,
together with \eqref{comHLA}, do not define an abstract Lie algebra by themselves.
To obtain a consistent set of algebra generators, one must also consider the following
 algebraic (crossing) relations among them \cite{fh15, fh22}:
\begin{align}
\label{crosLL}
&L_{ij}(L_{kl} - S_{kl})+L_{jk}(L_{il} -  S_{il})+L_{ki}(L_{jl} -  S_{jl})=0,
\\[1mm]
\label{crosLA}
 &L_{ij} A^\gamma_{k} + L_{jk} A^\gamma_i+ L_{ki} A^\gamma_j =0.
\end{align}
The relations \eqref{comLA} and \eqref{crosLA} remain valid upon replacing
$A^\gamma_i$ with $x_i$ and $\pi_i$.

The relations \eqref{comLL} and \eqref{crosLL}
define a quadratic algebra $\deform{so(N)}$, which forms a subalgebra
of the Cherednik algebra. The Dunkl angular momentum
provides  its representation \cite{fh15}; see also the recent work \cite{feigin26}.
The Casimir element of $\deform{so(N)}$ corresponds to
the angular part of the Hamiltonian \eqref{Hgamma}, denoted below by $H_\Omega$. It differs
from the angular momentum squared by an exchange-dependent term \cite{fh15}:
\begin{equation}
\label{CsoN}
H_\Omega = L^2+S(S-N+2),
\qquad
L^2=\sum_{i<j}L_{ij}^2,
\qquad
[H_\Omega,L_{ij}]=0.
\end{equation}
Note also that  the square of the vector $A^\gamma$
can be expressed in terms of the Casimir element and the Hamiltonian:
\begin{equation}
\label{Agamma-sq}
(A^\gamma)^2=\gamma^2 + 2H_\gamma \big( H_\Omega-S+\sfrac{(N-1)^2}{4}\big).
\end{equation}

One can extend $\deform{so(N)}$ by one additional dimension by incorporating
the deformed Laplace--Runge--Lenz invariant into the extra dimension
and trivially extending the exchange matrix:
\begin{equation}
\label{tLij}
\begin{aligned}
\tilde L_{ij}&=L_{ij}, \qquad \tilde S_{ij}=S_{ij}
\qquad
\text{for}
\quad i,j\le N,
\\
\tilde L_{i\,N+1}&=-\tilde L_{N+1\,i}=\frac{A^\gamma_i}{ \sqrt{-2 \Hgencoul}},
\qquad
\tilde L_{N+1\,N+1}=0,
\\
\tilde S_{i\,N+1}&=\tilde S_{N+1\,i}=0,
\qquad
\tilde S_{N+1\,N+1}=1.
\end{aligned}
\end{equation}
The extension  preserves the symmetries of the two tensors:
$\tilde S_{ij}=\tilde S_{ji}$, and $\tilde L_{ij}=- \tilde L_{ji}$.
The five relations \eqref{comLL}-\eqref{crosLA} can then
be expressed in the compact forms \cite{fh22}:
\begin{align}
\label{comLLt}
&[\tilde L_{ij},\tilde L_{kl}]=\imath \tilde L_{ik}\tilde S_{lj} + \imath \tilde L_{jl}\tilde S_{ki}
 - \imath \tilde L_{il}\tilde S_{kj} - \imath \tilde L_{jk}\tilde S_{li},
\\[1mm]
\label{crosLLt}
&\tilde L_{ij}(\tilde L_{kl} - \tilde S_{kl})+\tilde L_{jk}(\tilde L_{il} -  \tilde S_{il})
    +\tilde L_{ki}(\tilde L_{jl} -  \tilde S_{jl})=0.
\end{align}
So, the tilded operators form a particular representation of the algebra \eqref{comLL} and \eqref{crosLL} in $(N+1)$-dimensional space, namely the
$\deform{so(N+1)}$ algebra \cite{fh22}.

Its Casimir element, $\tilde H_\Omega$,  can be expressed in terms of the  Hamiltonian $H_\gamma$:
\begin{equation}
\label{CsoN+1}
 \tilde H_\Omega = \tilde{L}^2 + S(S-N+1)
=-\frac{\gamma^2}{2H_\gamma}-\frac{(N-1)^2}{4},
\qquad
[\tilde H_\Omega,\tilde L_{ij}]=0.
\end{equation}
Indeed, the first equality follows immediately from the expression for the angular Hamiltonian \eqref{CsoN} after the substitution $N\to N+1$.
This is a consequence of the fact that the formulas \eqref{tLij} define a representation of the
$\deform{so(N+1)}$ algebra. The second equality in \eqref{CsoN+1}
follows from Eqs.~\eqref{Agamma-sq} and \eqref{tLij}.
It relates the eigenvalues of the Casimir element to those of the Hamiltonian  \eqref{CsoN+1}
and can be inverted as follows:
\begin{equation}
\label{H-cas}
H_\gamma=-\frac{2}{\gamma^2}\left[\tilde H_\Omega+\frac{(N-1)^2}{4}\right]^{-1}.
\end{equation}
In the absence of the inverse-square interaction, this reduces to  the well-known relation between the
Coulomb Hamiltonian and the Casimir element of its $so(N+1)$ symmetry.

\section{Conformal algebra and the equidistant Calogero--Coulomb system}
\label{sec:conf}

\subsection{Three-dimensional conformal algebra}
First, we introduce the three-dimensional conformal group
$$
SL(2,\mathbb{R})\equiv SO(1,2)\equiv SU(1,1)
$$
which appears in the context
of the Calogero--Coulomb model. As in the standard hydrogen atom \cite{barut67,barut71,barut-book},
this group forms part of the dynamical symmetry of the system. In our case, it is generated
by the following operators:
\begin{gather}
K_1=\frac{r}{2}(\pi^2+1),
\qquad
K_2=\frac{r}{2}(\pi^2-1),
\qquad
K_3
=-\imath\left(r\partial_r+\sfrac{N-1}2\right),
\label{K123}
\\
 [K_\alpha,K_\beta]=-\imath\epsilon_{\alpha\beta\gamma}K^\gamma.
\label{comK}
\end{gather}
The restriction of these generators to symmetric wave functions, for which the Hamiltonian with particle exchange \eqref{Hgamma}
reduces to the conventional Calogero--Coulomb system \eqref{coul}, was considered in Ref.~\cite{khare99}.
Here, $\epsilon_{\alpha\beta\gamma}$ represents the Levi--Civita tensor in three dimensions, and the index is raised using the Minkowski metric $\eta_{\alpha\beta}$ with signature $(+--)$. The commutation relations \eqref{comK} become transparent once we note that the $g$-dependent term in the Dunkl momentum cancels out owing to the relation $[\pi_i,r]=[p_i,r]=-\imath \frac{x_i}{r}$.
The  Casimir element of the resulting Lie algebra  is then given by
\begin{equation}
\label{Ksq}
K^2=K_\alpha K^\alpha =K_1^2-K_2^2-K_3^2,
\qquad
[K^2,K_\alpha]=0.
\end{equation}
The conformal algebra mainly reflects the algebraic properties associated with the radial coordinates.
In particular, the generators \eqref{K123} commute with the Dunkl angular momentum operators, since
 $\pi^2$ commutes with them \cite{fh15}:
\begin{equation}
\label{comLK}
[L_{ij},K_\alpha]=0.
\end{equation}
In other words, they are $\deform{so(N)}$ scalars.
The operator $K^2$ also appears in the representation of the conformal generators
in terms of the radial components:
\begin{align}
\label{Krad}
& K_1=\frac{r}{2}(p_r^2+1)+\frac{K^2}{2r},
\qquad
K_2=\frac{r}{2}(p_r^2-1)+\frac{K^2}{2r},
\qquad
K_3=r p_r
\\
\label{pr}
& \text{with}
\quad
p_r =r^{-\frac{N-1}2}(-\imath \partial_r)r^{\frac{N-1}2}=-\imath\left(\partial_r+\frac{N-1}{2r}\right).
\end{align}
Recall that the definition of the radial momentum   \eqref{pr} ensures the Hermicity:
$p_r^+=p_r$ \cite{thyssen-book}.
Using  the last equality in \eqref{Krad}  and
canonical commutation relation $[r,p_r]=\imath$,   one obtains
the explicit expression for the Casimir element of the conformal algebra \eqref{Ksq}:
\begin{equation}
\label{Ksq-pr}
K^2=(K_1-K_2)(K_1+K_2)-\imath K_3-K_3^2=r^2(\pi^2-p_r^2).
\end{equation}
Substituting this result into $K_1$ and $K_2$  in  \eqref{Krad} reproduces
their original definitions \eqref{K123}.

Note that the conformal generators \eqref{K123} are not Hermitian with respect to the standard
measure due to the explicit factor of $r$ in their definitions. They become Hermitian
with respect to a modified measure that removes this factor:
\begin{equation}
\label{measure}
\langle\psi_1|\psi_2\rangle = \int r^{-1} \psi_1^*(x) \psi_2(x) d^N x= \int  \psi_1^*\psi_2 r^{N-2} dr d^{N-1} \Omega,
\end{equation}
where the second integral is written in spherical coordinates, and $\Omega$ denotes the solid angle.

The first one is a result of comparison of  \eqref{K123} and \eqref{Krad}.
 The second is a definition of the angular Hamiltonian of the Calogero model
 with particle exchange \eqref{CsoN}.

 Consider now two expressions for the square of the Dunkl momentum.
The first follows from comparing Eqs.~\eqref{K123} and \eqref{Krad}.
The second arises from the definition of the angular Calogero
Hamiltonian  \eqref{CsoN} \cite{fh15}:
\begin{equation}
\label{pisq}
\pi^2=p_r^2+\frac{K^2}{r^2},
\qquad
\pi^2=-\partial_r^2-\frac{N-1}{r}\partial_r+\frac{H_\Omega}{r^2}.
\end{equation}
Comparing these two expressions and using \eqref{pr}, we obtain an explicit relation between the
Casimir elements of the $so(1,2)$ and $\deform{so(N)}$ algebras:
\begin{equation}
K^2=H_\Omega+\sfrac14 (N-1)(N-3).
\label{Ccnf-osN}
\end{equation}
Note that these two operators coincide in three-dimensional case,  $N=3$.
Relation \eqref{CsoN} then implies:
\begin{align}
L^2=K^2-\big(S-\sfrac{N-1}2\big)\big(S-\sfrac{N-3}2\big).
\label{LKsq}
\end{align}

\subsection{Equidistant analogue of Calogero--Coulomb Hamitonian}
\label{sub:Theta}
For the Calogero model (with an oscillator potential instead of the Coulomb one),
the conformal algebra is generated by ladder operators \cite{isakov,hak23}
that separately produce the even and odd parts of the spectrum.
The spectrum of the Calogero--Coulomb Hamiltonian, however, is not equidistant;
see Eq.~\eqref{En} below. As a result, the overall picture becomes more involved.

As mentioned in the Introduction, the Hamiltonian of the hydrogen atom can be transformed into
an equivalent one with the same wave functions but an equidistant energy spectrum.
This property makes it possible to construct a spectrum-generating algebra
for the transformed Hamiltonian \cite{barut67,barut71};
see also \cite{barut-book,thyssen-book} for reviews.
Here, we generalize this construction,
originally applied to the hydrogen atom, to the Calogero--Coulomb system
with particle exchange \eqref{Hgamma}.

Multiplying the  Schr\"odinger equation
\begin{equation}
\label{schro}
H_\gamma\Psi=E\Psi,
\end{equation}
by $r$, we obtain an equivalent equation for the same wave function:
\begin{align}
\label{ThetaPsi}
\Theta_E\Psi&=\gamma\Psi,
\\
\label{Theta}
\Theta_E
&=\ \frac{r \pi^2}2-Er=\frac{rp^2}{2}+\sum_{i<j}\frac{g(g-s_{ij})r}{2(x_i-x_j)^2}-Er.
\end{align}
In the new Schr\"odinger equation \eqref{ThetaPsi},
the  energy $E$ and Coulomb coupling $\gamma$ interchange their roles:
the former becomes a parameter of the Hamiltonian,
whereas the latter appears as the eigenvalue.

A crucial property of the new Hamiltonian is that it can be expressed as a
linear combination of the conformal group generators \eqref{K123}:
\begin{equation}
\Theta_E=\frac12(1-2E)K_1+\frac12(1+2E)K_2.
\end{equation}
The commutation relations \eqref{comK} imply then that this element is, up to a
multiplicative factor, equivalent to the first generator:
\begin{equation}
\label{ThetaK}
\Theta_E=\sqrt{-2E}\,e^{-\imath\beta K_3} K_1 e^{\imath\beta K_3}
\qquad
\text{with}
\qquad
\tanh\beta=\frac{1+2E}{1-2E}.
\end{equation}
Later, we apply the above $\beta$-rotation to the entire conformal algebra
in order to align the generator $K_1$ with the Hamiltonian $\Theta_E$;
see Eqs.~\eqref{bKA} and \eqref{ThetaK'} below.
For $\beta=0$ the  constructed Hamiltonian  \eqref{Theta} coincides with $K_1$:
\begin{equation}
\label{ThetaK1}
\Theta_{-\frac12}=K_1.
\end{equation}
Evidently, $\Theta_E^\dag=\Theta_E$ since $K_\alpha^\dag=K_\alpha$,
where the dagger ($\dag$) denotes the Hermitian conjugate with respect to the
modified inner product \eqref{measure}.

Note that  the Calogero--Coulomb Hamiltonians \eqref{Hgamma} and \eqref{Theta}
admit a simple representation in terms of the Casimir element
of the $so(1,2)$ algebra \eqref{Ksq}:
\begin{align}
\label{HKsq}
H_\gamma&=\frac12 p_r^2+\frac{K^2}{2r^2}-\frac{\gamma}{r},
\\[2mm]
\label{TKsq}
\Theta_E&=\frac12 r p_r^2+\frac{K^2}{2r}-Er.
\end{align}
These expressions are analogous to the representation \eqref{Krad} of the conformal
generators and follow directly from Eq.~\eqref{Ksq-pr}.

Finally, we mention that the conformal algebra participates in  the dynamical symmetry
 and underlies the superintegrability of the rational Calogero models
 \cite{woj83,kuznetsov,gonera}.
 It also generates new invariants \cite{hak-kriv,hak14-2} for more general
 class of conformal mechanical systems \cite{fubini}.

\section{Dynamical symmetry of the  equidistant Calogero--Coulomb system}
\label{sec:dyn}

In this section, we construct
the complete spectrum-generating algebra for the equidistant counterpart of
the Calogero--Coulomb system \eqref{Theta}. Owing to the similarity
relation \eqref{ThetaK}, it suffices to first consider the special case
$E=-1/2$, for which the Hamiltonian reduces to the conformal-group generator $K_1$, see Eq.~\eqref{ThetaK1}.

\subsection{Dynamical symmetry algebra for the parameter $E=-1/2$}
The dynamical symmetry of the Calogero--Coulomb model \eqref{Hgamma}
consists of the Dunkl angular momentum algebra  $\deform{so(N)}$  \eqref{Lij},
the conformal $so(1,2)$ algebra \eqref{K123}, and three operator-valued Dunkl vectors
defined as follows:
\begin{align}
\label{AsigK}
A_{\sigma i}&=  \frac{x_i}{r}K_{\bar\sigma}- K_3 \pi_i
\quad\text{with}\quad \sigma=1,2,
\quad
\bar1=2,\quad \bar2=1,
\\
\Gamma_i&= r \pi_i.
\label{Gi}
\end{align}
All quantities above are obtained from the standard operators used in the context of the
hydrogen atom \cite{barut71,barut-book} by replacing the canonical momentum
with the Dunkl momentum, $p_i\to \pi_i$. A more conventional form can be obtained
by using the definitions \eqref{K123} and \eqref{Krad}:
\begin{align}
\label{Asig}
A_{\sigma}&=\frac12 x\left(\pi^2+(-1)^{\bar\sigma}\right)- rp_r \pi.
\end{align}
The vector $A_{1i}$ is an analogue of the Laplace--Runge--Lenz vector
associated with the modified Hamiltonian \eqref{ThetaK1}.
It differs from the corresponding vector \eqref{Agamma} defined for the original
Hamiltonian \eqref{Hgamma}.
In terms of the $\sigma$ notation introduced above, Eqs.~\eqref{comK} take the form
\begin{align}
\label{comKs}
[K_{\sigma_1},K_{\sigma_2}]=\imath \epsilon_{\sigma_1\sigma_2}K_3,
\qquad
[K_3,K_\sigma]=\imath K_{\bar\sigma}.
\end{align}

We start from the commutation relations  between the vectors \eqref{AsigK}, \eqref{Gi}
and the generators of the conformal group \eqref{K123}.
Since the latter are scalars and do not act on coordinate indices
\eqref{comLK}, we suppress the indices:
\begin{align}
\label{comKsAG}
[K_{\sigma_1},A_{\sigma_2}]&=-\imath \epsilon_{\sigma_1\sigma_2}\Gamma,
&
[K_\sigma,\Gamma]&=\imath A_{\bar\sigma},
\\[3pt]
\label{comK3AG}
 [K_3,A_\sigma]&=\imath A_{\bar\sigma},
&
[K_3,\Gamma]&=0.
\end{align}
Eqs.~\eqref{comKsAG} are proven in \ref{app:comm}; see Eqs.~\eqref{comKsA} and \eqref{comKsG} therein.
Eqs.~\eqref{comK3AG} follow immediately from the fact that the adjoint action of $\imath K_3$ measures the degree of a polynomial in the coordinates.
These four  relations can be combined into a single commutator:
\begin{equation}
\label{comKA}
[K_\alpha,A_\beta]=-\imath \epsilon_{\alpha\beta\gamma}A^\gamma,
\qquad
\text{where}
\qquad  A^3=\Gamma,
\end{equation}
and $\alpha,\beta,\gamma=1,2,3$. In other words, each component of the three vectors  $A_\beta$
forms the adjoint representation of the conformal algebra.
Nevertheless, taking into account
the structural difference between $A_\sigma$ and $\Gamma$, which will
certainly affect their algebraic properties, we prefer
to treat them separately.

Commutation relations between  the components of
$A_\sigma$  and $\Gamma$
 have a more complex form:
\begin{align}
\label{comAsAs}
{} [A_{\sigma_1 i},A_{\sigma_2 j}]&=\imath\epsilon_{\sigma_1\bar\sigma_2} L_{ij}+\imath\epsilon_{\sigma_1\sigma_2}K_3 S_{ij},
\\
\label{comGG}
{} [\Gamma_i,\Gamma_j]&=-\imath L_{ij},
\\
\label{comAsG}
{} [A_{\sigma i},\Gamma_j]&= \imath K_{\bar\sigma}S_{ij}.
&\quad&
\end{align}
They are derived in \ref{app:comm}; see  Eqs.~\eqref{comAA'}, \eqref{comGG'},
and \eqref{comAsG'} therein.
We see that the commutator of the components of the same vector
produces the Dunkl angular momentum element.
In contrast, the commutator between components of different vectors is
proportional to the product of the particle-exchange operator and a generator of the conformal algebra.

In addition, the commutator of any vector with the Dunkl angular momentum
assumes a unique form; see also Eq.~\eqref{comLA}:
\begin{equation}
\label{comLAs}
[L_{ij}, A_{k}]=   \imath A_{ i} S_{jk} -\imath A_{j} S_{ik}
\qquad
\text{with}
\qquad
A_k = A_{\alpha k}.
\end{equation}

The constructed generators of the dynamical symmetry
become Hermitian with respect to the modified measure \eqref{measure}.
Indeed, the additional factor of $r^{-1}$ in the integral does not affect on the Hermiticity  of
the  Dunkl angular momentum tensor.
On the other hand, the generators $r^{-1}K_\alpha$ and $r^{-1}\Gamma_i$ are Hermitian
in the standard measure, as follows from their definitions \eqref{K123}, \eqref{pr}, and \eqref{Gi}.
Therefore, $K_\alpha^\dag=K_\alpha$ and $\Gamma_i^\dag=\Gamma_i$.
Finally, the condition $A_{\sigma i}^\dag=A_{\sigma i}$ follows from the second commutators in
Eqs.~\eqref{comKsAG} and \eqref{comK3AG}.

We now define an antisymmetric matrix $\gen_{ab}=-\gen_{ba}$,
which unifies all generators of the dynamical symmetry:
\begin{equation}
\label{Lgen}
\begin{aligned}
\gen_{ij}=L_{ij},
\qquad
\gen_{i,N+1}=A_{1i},
\qquad
\gen_{i,N+2}=A_{2i}, 
\qquad
\gen_{i,N+3}=\Gamma_i, 
\\
\gen_{N+1,N+2}=K_3,
\qquad
\gen_{N+1,N+3}=K_2,
\qquad
\gen_{N+2,N+3}=K_1,
\end{aligned}
\end{equation}
where $1\le a,b\le N+3$ and $1\le i,j\le N$.
The commutation relations \eqref{comK}, \eqref{comLK}, \eqref{comKA}, \eqref{comAsAs}, \eqref{comGG}, \eqref{comAsG}, and \eqref{comLAs} can be compactly encoded then into the single relation:
\begin{equation}
\label{comLLgen}
[\gen_{ab},\gen_{cd}]=\imath(\gen_{bc}\cf_{ad}+\gen_{ad}\cf_{bc}-\gen_{bd}\cf_{ac}-\gen_{ac}\cf_{bd}).
\end{equation}
Here, the symmetric matrix $\cf_{ab}=\cf_{ba}$, which encodes the (deformed)
structure constants including particle-exchange effects,  has the following nontrivial
entries ($1\le i,j\le N$):
\begin{align}
\label{Sgen}
\cf_{ij}=S_{ij},
\qquad
\cf_{N+1N+1}=1,
\qquad
\cf_{N+2N+2}=\cf_{N+3N+3}=-1.
\end{align}
For clarity, let us present the matrix structure of  $\gen_{ab}$ and $\cf_{ab}$,
showing only the upper-triangular elements:
\begin{equation}
\label{genMat}
\gen=
\left(
\begin{array}{c  c  c c  c |  c  c }
0 & L_{12} & \cdots & L_{1N} &  A_{11} & A_{21} & \Gamma_1 \\
 & 0 & \cdots & L_{2N} &  A_{12} & A_{22} & \Gamma_2\\
  &   & \ddots & \vdots & \vdots & \vdots & \vdots  \\
 &  &   & 0 &  A_{1N} & A_{2N} & \Gamma_N\\[6pt]
 &  & &  &  0 & K_3 & K_2\\[3pt]
\hline
 &  &  &  &  &  0 & K_1\\
 &  & & &   & & 0\\
\end{array}
\right),
\qquad
\cf=
\left(
\begin{array}{c  c  c l | c   c  }
S_{11}  & \cdots & S_{1N} &  0\;\; & 0 & 0 \\
  &   \ddots & \vdots & \vdots & \vdots & \vdots  \\
 &   & S_{NN} &  0 & 0 & 0 \\
 &  & & 1 & 0 & 0 \\
\hline
 &  &  & &  -1 & 0\\
 & &  &   & & -1\\
\end{array}
\right).
\end{equation}
In both matrices, the lines separate blocks with different signatures: the diagonal
blocks have positive signature, while the off-diagonal blocks have negative signature.

In the $g=0$ limit, corresponding to the pure Coulomb model, the algebra
\eqref{comLLgen}, \eqref{Sgen} reduces to  the $so(N+1,2)$
Lie algebra, with structure constants determined by the pseudo-Euclidean metric
$\cf_{ab}=\epsilon_a\delta_{ab}$, whose signature is
$\epsilon_1=\dots=\epsilon_{N+1}=1$ and $\epsilon_{N+2}=\epsilon_{N+3}=-1$.
This algebra generates the conformal group in $(N+1)$-dimensional
Minkowski space.

For general $g$, the commutation relations \eqref{comLLgen} have a structure similar
to that of the angular momentum algebra \eqref{comLL}.
We therefore refer to the associative algebra generated by the operators
$\gen_{ab}$ and $\cf_{ab}$ as
 the full Dunkl conformal algebra, $\deform{so(N+1,2)}$.
This algebra contains, as subalgebras, the Dunkl angular momentum algebra $\deform{so(N)}$
 and the full symmetry algebra  $\deform{so(N+1)}$.
The generators also satisfy additional algebraic relations
analogous to the crossing relations \eqref{crosLL} and \eqref{crosLLt}.
A detailed investigation of these relations and the structure of the resulting algebra lies
beyond the scope of the present work.

Let us now derive the (quadratic) Casimir element of the $\deform{so(N+1,2)}$
algebra.
First, note that the squares of all three vectors, similarly to the Dunkl angular momentum squared
\eqref{LKsq}, can  be expressed in terms of the $SL(2,\mathbb{R})$ generators:
\begin{align}
A_1^2&= K_2^2+K_3^2 + S-\sfrac{N-1}{2},
\label{Asq}
\\
A_2^2&= K_1^2-K_3^2 - S+\sfrac{N-1}{2},
\label{Msq}
\\
\Gamma^2&=K_1^2 -K_2^2-S+\sfrac{N-1}2.
\label{Gsq}
\end{align}
The derivation of the above formulas is provided in \ref{app:Asq};
see Eqs.~\eqref{Asq'} and \eqref{Gsq'}.
Combining the above three equations we obtain:
\begin{equation}
\label{AalSq}
A_\alpha^2=K_\alpha^2+\eta_{\alpha\alpha}\left(K^2- S+\sfrac{N-1}{2}\right),
\end{equation}
where $\alpha=1,2,3$, and  $\eta_{\alpha\beta}$ is the Minkowski
metric associated with the three-dimensional conformal algebra;
see the text below Eq.~\eqref{comK}.

We now ready to evaluate the squared sum of all entries of the matrix \eqref{genMat},
with their signatures  taken into account. Using Eqs.~\eqref{LKsq} and
\eqref{AalSq}, we find:
\begin{equation}
\label{calLsq}
\begin{aligned}
\gen^2  &= \sum_{a<b}\gen_{ab}\gen^{ab}=L^2+A_1^2-A_2^2-\Gamma^2+K_1^2-K_2^2-K_3^2
\\
&=-S(S-N-1)-\frac14 {(N-1)(N+3)}.
\end{aligned}
\end{equation}
While the binomial in $S$ represents the quadratic invariant  of the permutation
group algebra
\eqref{S}, it does not commute with the elements $\gen_{ab}$.
Hence,  to construct the Casimir element, denoted by $\mathcal{H}_\Omega$,
 this term must be subtracted from $\gen^2$.
 In the  present representation, it reduces to a constant, yielding:
\begin{equation}
\label{cas-gen}
\mathcal{H}_\Omega=\gen^2+S(S-N-1)=-\frac14(N-1)(N+3).
\end{equation}
The Casimir operators $H_\Omega$, $\tilde H_\Omega$
and  $\mathcal{H}_\Omega$ of $\deform{so(N)}$, $\deform{so(N+1)}$,
and $\deform{so(N+1,2)}$, respectively, share the same structural form,
$$
\text{Casimir}=(\text{generators})^2+S(S-d+2),
$$
where $d$ is the dimension of the underlying space ($d=N,N+1,N+3$).
Note that recently a procedure for constructing the Casimir operators
of the $W_N$ algebra, which appears in the context of the Calogero model,
has been proposed \cite{correa23}.

\subsection{Dynamical symmetry for general $E$}
Consider now the barred  elements of the $\deform{so(N+1,2)}$ algebra,
 obtained from the original ones
via a $\beta$-rotation about the $K_3$ axis \eqref{ThetaK}:
\begin{equation}
\label{bgenL}
\bar\gen_{ab}=e^{-\imath\beta K_3} \gen_{ab} e^{\imath\beta K_3},
\qquad
\bar\cf_{ab}=\cf_{ab}.
\end{equation}
This transformation leaves the structure constants unchanged,
as the conformal algebra is invariant under particle exchange.
Certain components of the tensor $\gen_{ab}$  also remain invariant; see
Eqs.~\eqref{comLK} and \eqref{comKA}.
The first equation  in \eqref{bgenL}
can be expressed in expanded form as follows:
\begin{gather}
\label{bKA}
\bar K_\sigma = e^{-\imath\beta K_3} K_\sigma e^{\imath\beta K_3},
\qquad
\bar A_{\sigma i } = e^{-\imath\beta K_3} A_{\sigma i} e^{\imath\beta K_3},
\\
\bar L_{ij} = L_{ij},
\qquad
\bar K_3 =K_3,
\qquad
\bar A_{3i} =A_{3i}.
\end{gather}
where, as before, $\sigma=1,2$.
Clearly, the transformed $\deform{so(N+1,2)}$ generators satisfy
the same commutation relations as the
original ones \eqref{comLLgen}:
\begin{equation}
\label{comLLgen'}
[\bar\gen_{ab},\bar\gen_{cd}]=\imath(\bar\gen_{bc}\cf_{ad}+\bar\gen_{ad}\cf_{bc}-\bar\gen_{bd}\cf_{ac}-\bar\gen_{ac}\cf_{bd}).
\end{equation}
The squares of the conformal algebra \eqref{Ksq} and of the dynamical symmetry algebra \eqref{calLsq}
remain invariant under the $\beta$-rotation:
\begin{equation}
\bar K^2=K^2,
\qquad
\bar{\mathcal{L}}^2=\mathcal{L}^2.
\end{equation}

Using the definition of the hyperbolic angle $\beta$ \eqref{ThetaK} and the commutation relations
\eqref{comKs}, \eqref{comK3AG}, one can straightforwardly derive the explicit forms of the rotated
generators \eqref{bKA}:
\begin{align}
\label{bK}
\bar K_\sigma&=\frac{r}{\sqrt{-2E}}\left(\frac{\pi^2}{2}+(-1)^{\sigma}E \right),
\\
\label{bA}
\bar A_\sigma&=\frac{1}{\sqrt{-2E}}\left[x\left(\frac{\pi^2}2+(-1)^{\sigma}E \right) +rp_r\pi\right].
\end{align}
Equations \eqref{bK} indicate that the equidistant form of the Calogero--Coulomb Hamiltonian
\eqref{Theta}  is proportional to the shifted   generator corresponding to $\sigma=1$:
\begin{equation}
\label{ThetaK'}
\Theta_E=\sqrt{-2E}\bar K_1.
\end{equation}
This relation determines  the
spectrum of  $\Theta_E$, since  the conformal element $K_1$
in the discrete representation assumes positive   integer eigenvalues.

The Dunkl angular momentum  tensor and the $\beta$-rotated modified Runge-Lenz  vector $\bar A_1$
generate the full symmetries of the above Hamiltonian:
\begin{equation}
\label{symTheta}
[\Theta_E,L_{ij}]=[\Theta_E, \bar A_{1i}]=0.
\end{equation}
The latter  depends on $E$
and  represents an additional integral of motion,
analogous to the invariant \eqref{Agamma}  for the original
Calogero--Coulomb Hamiltonian.

At the same time, the scalars $\bar K_2$, $K_3$ together with the vectors $\bar A_2$
and  $A_3=-\Gamma$ form the spectrum generated sector of the $\deform{so(N+1,2)}$ algebra.
The associated  ladder operators are defined by:
\begin{align}
\label{Kpm'}
\bar  K_\pm &= \bar K_2\pm \imath K_3:
\qquad
&
[\bar K_1,\bar K_\pm]&=\pm\bar K_\pm,
\quad
&
[\bar K_+,\bar K_-]&=-2\bar K_1,
\\[2pt]
\label{Apm'}
\bar A_{\pm i} &= \bar A_{2i}\pm \imath A_{3i}:
\qquad
&
[\bar K_1,\bar A_{\pm i}]&=\pm \bar A_{\pm i},
\quad
&
[\bar A_{+i},\bar A_{-i}]&=-2\bar K_1S_{ii}.
\end{align}
The above commutation relations follow from Eqs.~\eqref{comK} and \eqref{comKA}.
The correspondence  \eqref{ThetaK'} implies that the ladder operators $\bar K_\pm$ and $\bar A_{\pm i}$
map between two energy levels of the equidistant Calogero--Coulomb
Hamiltonian \eqref{Theta} separated by $\sqrt{-2E}$:
\begin{align}
\label{comThetaKpm}
[\Theta_E,\bar K_{\pm}]&=\pm \sqrt{-2E} \bar K_{\pm},
\\
\label{comThetaApm}
[\Theta_E,\bar A_{\pm i}]&=\pm \sqrt{-2E} \bar A_{\pm i}.
\end{align}

\section{$SO(1,2)$ structure of the Calogero--Coulomb wave functions}
\label{sec:conf-struc}

In this section, we first construct explicitly the eigenfunctions of the equidistant
Calogero--Coulomb Hamiltonian in terms of the deformed spherical harmonics.
We then analyze the representations of the three-dimensional conformal algebra on these states.

\subsection{Stationary wave functions in deformed spherical harmonics}
Recall the Maxwell representation of the deformed homogeneous  harmonic polynomials \cite{Xu-2000} (see also
\cite{hak23}). These polynomials are labeled by the magnetic quantum numbers $m_i=0,1,2,\dots$:
\begin{align}
\label{hn}
h_{m_1\dots m_N}(x)&=r^{2(c_g-1+l)+N}
\nabla'^{m_1}_1\dots \nabla'^{m_N}_N
r^{-2(c_g-1)-N}.
\end{align}
Here $l$ denotes the  degree of the polynomial associated
with the angular momentum (orbital) quantum number,
while the constant $c_g$  accounts for the Calogero interaction:
\begin{align}
\label{l}
l&=\sum_{i=1}^N m_i,
\\
\label{cg}
c_g&=\sfrac12 gN(N-1).
\end{align}
The primed nablas in Eq.~\eqref{hn} are obtained from the standard Dunkl operators
through a (nonunitary) similarity transformation:
\begin{align}
\label{du'}
\nabla'_i&=\phi^{-1}\nabla_i\phi=\partial_i+\sum_{j\ne i}\frac{g}{x_i-x_j}(1-s_{ij}),
\\
 \label{phi}
\phi(x)&=\prod_{i<j}|x_i-x_j|^g.
\end{align}
The solution of the Shr\"odinger equation for  the angular Calogero model \eqref{CsoN} is
given then by \cite{flp}:
\begin{align}
\label{phihn}
H_\Omega\, \phi(x) h_{m_1\dots m_N}(x)&=\mathcal{E}_l\phi(x) h_{m_1\dots m_N}(x),
\\
\label{epsilon}
{\cal E}_l&=(c_g+l)(c_g+l+N-2).
\end{align}
The energy levels depend solely on the angular
momentum (orbital) quantum number $l$.
This property leads to the highest level of degeneracy
and ensures the maximal superintegrability of the angular system \cite{flp,hak14-1}.

The wave functions in Eq.~\eqref{phihn} are unnormalized and satisfy a linear dependence condition;
see Eq.~\eqref{dep} below. They form an overcomplete basis,
and the orthogonality condition holds only for different values
of the orbital quantum number $l$ \cite{Xu-2000}.

The eigenfunctions of the Calogero--Coulomb Hamiltonian \eqref{Hgamma}
can then be expressed in spherical coordinates as:
\begin{align}
\label{psik}
\Psi_{m_1\dots m_N,k}(x)&= \text{const}\cdot e^{-\frac{\rho}{2}}L_k^{(2\alpha_l)}(\rho)\phi(x) h_{m_1\dots m_N}(x),
\\
\label{Hpsik}
H_\gamma \Psi_{m_1\dots m_N,k} &= E_{n}\Psi_{m_1\dots m_N,k}.
\end{align}
Here, $L_k^{(\alpha)}$ is the associated Laguerre polynomial, where the lower
index $k$ corresponds to the
radial quantum number, counting the nodes of the  wave function.
The upper index $\alpha$ is determined by the orbital quantum number,
while the energy depends on the principal quantum number:
\begin{align}
\label{alpha}
\alpha_l&=c_g+l+\sfrac12 N-1,
\\
\label{n}
 n&=l+k+1,
 \qquad
 k=n_r=0,1,2,\dots,
 \\
\label{En}
E_n&=-\frac{\gamma^2}{2(c_g+n+\sfrac{N-3}2)^2}.
\end{align}
Finally, the argument of the Laguerre polynomial is the renormalized, dimensionless radial coordinate:
\begin{equation}
\label{rho}
\rho=\rho_n=\frac{2\gamma r}{c_g+n+\frac{N-3}2}.
\end{equation}
The Shr\"odinger equation  \eqref{Hpsik} may be  verified by substituting \eqref{Ccnf-osN} into the
radial form of the Hamiltonian \eqref{HKsq} and using the angular
Shr\"odinger equation  \eqref{phihn}, \eqref{epsilon}.

The last two terms in the wave function \eqref{psik} are homogeneous polynomials of
total degree $c_g+l$.
Therefore, they depend on the angular variables $u_i=x_i/r$, which completes the
separation of radial and angular coordinates:
\begin{equation}
\label{psik'}
\Psi_{m_1\dots m_N,k}(x)= \text{const}\cdot \rho^{c_g+l}e^{-\frac{\rho}{2}}L_k^{(2\alpha_l)}(\rho)\phi(u) h_{m_1\dots m_N}(u).
\end{equation}
%
%

According to the discussion in Sect.~\ref{sub:Theta}, the constructed wave function \eqref{psik'}
solves also the Shr\"odinger equation for the equidistant Calogero--Coulomb model \eqref{ThetaPsi}, \eqref{Theta}:
\begin{align}
\label{Theta-psi}
\Theta_E\Psi_{m_1\dots m_N, k} & = \gamma_n\Psi_{m_1\dots m_N, k},
\\
\label{gn}
\gamma_n& =\sqrt{-2E}\left(c_g+n+\sfrac{N-3}2\right),
\end{align}
where $n$ is the principal quantum number \eqref{En}.
Here, the energy value $E<0$ is considered as a parameter, and the  radial variable
\eqref{rho} must be reorganized though the energy as:
\begin{align}
\label{rhoE}
\rho=2\sqrt{-2E}\, r.
\end{align}
In contrast to the original spectrum \eqref{En}, the current spectrum \eqref{gn} is equidistant.
This property facilitates the construction of a spectrum-generating algebra, which produces
the complete set of energy eigenstates together with their spectrum from the ground state.
Note that the quantization conditions \eqref{En} and \eqref{gn} are equivalent
since both them are obtained from the relation  $\gamma/\sqrt{-2E}=c_g+n+\sfrac{N-3}2$.

\medskip

Upon symmetrization over all coordinates, the deformed harmonic polynomials
\eqref{hn} take the form \cite{flp}:
\begin{align}
\label{hk}
h_{k_1k_3\dots k_N}^\text{sym}(x)&=r^{2(c_g-1+l)+N}
\mathcal{D}_1^{k_1}\mathcal{D}_3^{k_3}\dots \mathcal{D}_N^{k_N}
r^{-2(c_g-1)-N},
\\
\label{lsym}
\mathcal{D}_q&=\sum_{i=1}^N \nabla'_i\mbox{}^q,
\qquad
l=k_1+\sum_{i=3  }^N i k_i.
\end{align}
Here the integers $k_i$ label the modes of a symmetric polynomial, while  $l$ detones its degree,
and corresponds to the angular momentum quantum number.  The Dunkl Laplacian $\mathcal{D}_2$
does not appear, since its action annihilates the power function
of the right-hand side of the above expression, $\mathcal{D}_2 r^{-2(c_g-1)-N}=0$.
As a consequence, the deformed harmonic
polynomials \eqref{hn} are linearly dependent  \cite{Xu-2000}:
\begin{equation}
\label{dep}
\sum_{i=1}^N h_{m_1\dots m_{i-1} (m_i+2)m_{i+1}\dots m_N}(x)=0.
\end{equation}

The eigenfunctions of the angular Hamiltonian
associated with the bosonic Calogero-Moser model $\tilde H_0$ \eqref{coul},
take the form \cite{flp}
\begin{equation}
\label{phihns}
H_\Omega\, \phi(x)h^\text{sym}_{k_1k_3\dots k_N}(x)=\mathcal{E}_l\phi(x) h^\text{sym}_{k_1k_3\dots k_N}(x).
\end{equation}
The corresponding energy is given by Eq.~\eqref{epsilon}, where the orbital quantum number
is determined by Eq.~\eqref{lsym}.

As a consequence, the solution of the Schr\"odinger
equation  of the standard Calogero--Coulomb model describing identical bosons
\eqref{coul} takes the form:
\begin{align}
\label{psik'sym}
\Psi^\text{sym}_{k_1k_2k_3\dots k_N}(x)&= \text{const}\cdot \rho^{c_g+l}e^{-\frac{\rho}{2}}L_{k_2}^{(2\alpha_l)}(\rho)\phi(u)
h_{k_1k_3\dots k_N}^\text{sym}(u),
\\
\label{Hpsiksym}
H_\gamma \Psi^\text{sym}_{k_1\dots k_N} &= E_{n}\Psi^\text{sym}_{k_1\dots k_N},
\\
\label{Thpsiksym}
\Theta_E\Psi^\text{sym}_{k_1\dots k_N} & = \gamma_n\Psi^\text{sym}_{k_1\dots k_N}.
\end{align}
The energy is given by Eq.~\eqref{En}, and the principal quantum number
is defined in Eq.~\eqref{n}.
The radial quantum number appears as the second quantum number
of the solution \eqref{psik'sym}: $n_r=k=k_2$.
Substituting  the orbital quantum number \eqref{lsym}
into  Eq.~\eqref{n}, we arrive at the principal quantum number
in the symmetric representation,
\begin{equation}
\label{nsym}
n=l+k_1+k_2+\sum_{i=3}^N ik_i+1.
\end{equation}
Symmetrization projects the operators onto their permutation-invariant counterparts.
Consequently, the rational Cherednik algebra reduces to the permuta\-ti\-on-invariant sector,
commonly referred to in the mathematical literature as its spherical subalgebra \cite{etingof}.

\subsection{Conformal algebra structure of eigenfunctions} 
The spectrum-generating sector of the full dynamical symmetry $\deform{so(N+1,2)}$ connects
different stationary states \eqref{psik'} of the Calogero--Coulomb Hamiltonians \eqref{Hgamma} an \eqref{Theta}.
In this section, we study the structure of states  connected by the three-dimensional conformal
subalgebra, $sl(2,\mathbb{R})\equiv so(1,2)$, which is invariant under particle exchange
and therefore acts consistently on the bosonic (fermionic) states \eqref{psik'sym}.

Substituting \eqref{pisq} into \eqref{bK} separates the radial and angular variables in $\bar K_\alpha$.
After introducing the renormalized radial coordinate \eqref{rhoE}, the resulting expressions
take a form similar to those obtained in the $E=-1/2$ case \eqref{Krad}:

\begin{align}
\label{Krho}
& \bar K_\sigma=\rho\left(p_\rho^2+\frac{K^2}{\rho^2}+\frac{(-1)^{\bar\sigma}}{4}\right),
\qquad
\bar K_3=\rho\, p_\rho.
\end{align}
Taking into account the relation \eqref{Ccnf-osN} between the Casimir elements of the conformal and
Dunkl angular momentum algebras, and using  Eqs.~\eqref{phihn}, \eqref{epsilon},
we get:
\begin{equation}
\label{Ksq-hn}
K^2\, \phi(u) h_{m_1\dots m_N}(u)=\left(\alpha_l^2-\sfrac14\right)\phi(u) h_{m_1\dots m_N}(u).
\end{equation}
Substituting this expression into $\bar K_\sigma$ \eqref{Krho}  and acting
on the stationary states yields:
\begin{align}
\label{K-psi}
\bar K_-\Psi_{m_1\dots m_N, k} &=(2\alpha_l+k)\Psi_{m_1\dots m_N, k-1},
\\
\label{K+psi}
\bar K_+\Psi_{m_1\dots m_N, k-1} &=k\Psi_{m_1\dots m_N, k}.
\end{align}
Here, the normalization constant   from the wave function \eqref{psik'} is omitted for simplicity,
and $l$ and $\alpha_l$ are defined in Eqs.~\eqref{l} and \eqref{alpha}, respectively.
Finally, the Schr\"odinger equation  \eqref{Theta-psi}, \eqref{gn} is equivalent to the
eigenvalue equation of the  generator $\bar K_1$ \eqref{alpha}:
\begin{equation}
\label{K1psi}
\bar K_1\Psi_{m_1\dots m_N, k}  =\left(\alpha_l+k+\sfrac{1}2\right)\Psi_{m_1\dots m_N, k}.
\end{equation}
Thus, for fixed magnetic quantum numbers $m_i$, the (unnormalized)
states
\\
$\Psi_{m_1\dots m_N, k}$
with $k=0,1,\dots$
form an infinite-dimensional
lowest-weight irreducible representation of the Lie algebra  $sl(2,\mathbb{R})$.
The conformal spin  and its square are given, respectively, by
\begin{equation}
\label{s}
s=\alpha_l+\sfrac12=c_g+l+\sfrac12(N-1),
\qquad
K^2=s(s-1).
\end{equation}
The lowest-weight  state is
\begin{equation}
\label{psi0}
\Psi_{m_1\dots m_N, 0}=\rho^{c_g+l}e^{-\frac{\rho}{2}}\phi(u) h_{m_1\dots m_N}(u).
\end{equation}

Clearly, the  multiplets characterized by different sets of magnetic quantum numbers $m_i$ are
equivalent provided that they have the same conformal spin $s$ and
the same orbital quantum number $l$.

The tower of states $\Psi_{m_1\dots m_N, k}$ has energy
levels given by Eqs.~\eqref{En} and \eqref{gn}, corresponding to the Calogero--Coulomb Hamiltonian
\eqref{Hgamma} and its equidistant counterpart \eqref{Theta}, respectively.
Both are  specified by the principal quantum number \eqref{n}, which can also be written
as follows:
\begin{equation}
\label{n-s}
n=s+k-c_g-\sfrac12(N-3).
\end{equation}

The constructed representation is unitary with respect to the measure
\eqref{measure}, with Hermitian generators satisfying:
$\bar K_-^\dag=\bar K_+$ and $\bar K_1^\dag=\bar K_1$.
As noted in the previous subsection, the deformed spherical harmonics \eqref{hn} are orthogonal only for distinct
orbital quantum numbers.  Consequently, the wave functions \eqref{psik}
are generally nonorthogonal,  but  remain orthogonal  for different orbital \eqref{l}
and radial quantum numbers:
\begin{equation}
\langle \Psi_{m_1\dots m_N, k}|  \Psi_{m'_1\dots m'_N, k'}\rangle
=
\text{const}\cdot\delta_{ll'}\delta_{kk'},
\end{equation}
where the constant depends on magnetic quantum numbers.

To verify the representation \eqref{K-psi}--\eqref{K1psi},
we apply the corresponding operators directly
to the Laguerre polynomials appearing in the wave function \eqref{psik'}.
Straightforward  calculations show  that these equations
are equivalent, respectively, to (with $a=2\alpha_l$):
\begin{align}
\label{K+L}
\left[\rho\frac{d^2}{d\rho^2}-(2\rho-a-1)\frac{d}{d\rho}+\rho-a-1\right] L_{k-1}^{(a)}(\rho)+kL_k^{(a)}(\rho)&=0,
\\
\label{K-L}
\left[\rho\frac{d^2}{d\rho^2}+(a+1)\frac{d}{d\rho}\right] L_{k}^{(a)}(\rho)+(a+k)L_{k-1}^{(a)}(\rho)&=0,
\\
\label{K1L}
\left[\rho\frac{d^2}{d\rho^2}-(\rho-a-1)\frac{d}{d\rho}+k\right] L_k^{(a)}(\rho)&=0,
\end{align}
which can be verified using the Rodrigues formula
for the associated Laguerre polynomials given by
\begin{equation}
L_k^{(a)}(z)=\frac1{k!}z^{-a}e^z\frac{d^k}{dz^k}\left(e^{-z}z^{k+a}\right).
\end{equation}

\medskip

As noted above, due to the permutation symmetry of the conformal algebra \eqref{K123},
its action can be projected onto an irreducible representation of the symmetric group.
In particular, the symmetrization procedure reduces the constructed multiplet
to an equivalent one with the same conformal spin.
The resulting representation is given on the bosonic wave functions \eqref{psik'sym}
and is structured in the same way:
\begin{align}
\label{K'-psi}
\bar K_-\Psi^\text{sym}_{k_1 k_2k_3\dots k_N  } &=(2\alpha_l+k_2)\Psi^\text{sym}_{k_1 (k_2-1)k_3\dots k_N },
\\
\label{K'+psi}
\bar K_+\Psi^\text{sym}_{k_1 (k_2-1)k_3\dots k_N  } &=k_2\Psi^\text{sym}_{k_1 k_2k_3\dots k_N  },
\\
\label{K1'psi}
\bar K_1\Psi^\text{sym}_{k_1 \dots k_N } &
=\left(\alpha_l+k_2+\sfrac{1}2\right)\Psi^\text{sym}_{k_1\dots k_N }
\end{align}
with the orbital quantum number $l$ defined by Eq.~\eqref{lsym}.
The conformal group again governs the radial quantum number $k_2$,
while the other modes $k_i$ ($i\ne2$) remain unaffected.

The lowest state is given by the $k_2=0$ wave function
$$
\Psi^\text{sym}_{k_10k_3\dots k_N}=\rho^{c_g+l}e^{-\frac{\rho}{2}}\phi(u) h^\text{sym}_{k_1k_3\dots k_N}(u).
$$
The conformal spin,  Casimir element, and the principal quantum number are given by the same formulas
as in Eqs.~\eqref{s} and \eqref{n-s}.

\section{Conclusion}
\label{sec:concl}
In this article, we investigate the dynamical symmetry of the
Calogero model in an external confining Coulomb field  within the framework of
 the exchange operator formalism.
Following the case of the interaction-free  Coulomb
model, we construct an alternative system that retains the same stationary
states while exhibiting  a linear (equidistant) spectrum, thereby admitting ladder operators.
The resulting Hamiltonian is identified as an element of the conformal algebra $so(1,2)$,
which underlies the dynamical symmetry of the model.

It was shown that the equidistant Calogero--Coulomb  Hamiltonian admits a Dunkl-deformed  Laplace--Runge--Lenz vector, which
differs from the analogous vector previously obtained for the original system \cite{runge}.
Together with the Dunkl angular momentum tensor, it
forms a complete symmetry algebra of the  equidistant  Calogero--Coulomb model,
 $\deform{so(N+1)}$. The latter has a structure similar to that of the original system \cite{fh22}.

Next, the $so(1,2)$ algebra extends the Dunkl-deformed orthogonal
symmetry to the full dynamical symmetry algebra, which can be regarded
as an exchange-operator deformation of the conformal algebra in $(N+1)$-dimensional
Minkowski space, $\deform{so(N+1,2)}$.
 The commutation relations between the generators have been
 derived explicitly. The corresponding structure constants generally
 depend on particle exchange.
 Nevertheless, all commutation relations can be compactly encoded in a single matrix expression
 involving the algebra generators and structure constants.
 Furthermore, ladder operators that shift the energy by one unit are constructed.
 The quadratic Casimir operator of the resulting deformed algebra is shown to
 be to a constant.

 Finally, we  construct the stationary states of the  Calogero--Coulomb
 Hamiltonian, both with and without  particle exchange.
 The construction relies on deformed spherical harmonics and their
 symmetrized counterparts \cite{dunkl,Xu-2000}.  We show that the
 wavefunctions are classified according to the infinite-dimensional
 lowest-weight irreducible representation of the $so(1,2)$ conformal algebra. The
 diagonal conformal generator associated to the "time" component is
 given by the (rescaled) equidistant Calogero--Coulomb Hamiltonian. The  conformal spin
 of this multiplet is determined by the Calogero coupling constant and orbital quantum
 number, while the states are labeled by the radial quantum number.

We emphasize that the commutation relations \eqref{comLLgen} and \eqref{comLLgen'}
do not define an abstract Lie algebra. To satisfy the Jacobi identity,
the  generators must obey additional constraints,
analogous to the crossing relations \eqref{crosLL},
whose derivation would also be of interest.

In this work, we have focused on the standard Calogero--Coulomb Hamiltonian,
which is invariant under particle permutations.
The present construction is likely to extend to systems with arbitrary finite-reflection group (Coxeter) symmetry.
The noninteracting Hamiltonians, in which the group is generated by independent reflections, have already been discussed in the literature; see Ref.~\cite{herranz25} and references therein.

It is worth noting that the spectrum-generating operators of the standard Calogero--Coulomb model \eqref{coul} can be expressed
as symmetric polynomials in the generators of the deformed conformal algebra.
The symmetrization procedure projects these operators onto their permutation-invariant counterparts,
thereby ensuring consistency with the underlying particle-exchange symmetry of the model.
However, this projection substantially complicates the algebraic relations among the resulting elements.

\section*{Acknowledgments}
The author thanks   Maria Matushko and Ivan Sechin  for valuable discussions.
This work was supported  by  Armenian Science Committee Grants Nos. 21AG-1C047 and
24FP-1F039.

\bibliography{dynCoul-refs-final}

@string{JMP = "J. Math. Phys."}

@string{PRL = "Phys. Rev. Lett."}

@string{PRA = "Phys. Rev. A"}

@string{PRD = "Phys. Rev. D"}

@string{PRev = "Phys. Rev."}

@string{PRep = "Phys. Rep."}

@string{PLA = "Phys. Lett. A"}

@string{PLB = "Phys. Lett. B"}

@string{JPA = "J. Phys. A"}

@string{LMP = "Lett. Math. Phys."}

@string{NPB = "Nucl. Phys. B"}

@string{IM = "Inv. Math."}

@string{JHEP = "JHEP"}

@string{TAMS = "Trans. Amer. Math. Soc."}

@string{PhysD = "Physica D"}

@string{AM = "Adv. Math."}

@string{CMB = "Canad. Math. Bull."}

@string{JPAA = "JPAA"}

@string{CMP = "Comm. Math. Phys."}

@article{calogero-1,
  author       = {Calogero, F.},
  title        = {Solution of a three-body problem in one dimension},
  journal      = JMP,
  volume       = {10},
  year         = {1969},
  pages        = {2191},
  doi          = {10.1063/1.1664820}
}

@article{calogero-2,
  author       = {Calogero, F.},
  title        = {{Solution of the one-dimensional $N$-body problems with quadratic and/or inversely quadratic pair potentials}},
  journal      = JMP,
  volume       = {12},
  year         = {1971},
  pages        = {419},
  doi          = {10.1063/1.1665604}
}

@article{moser,
  author       = {Moser, J.},
  title        = {Three integrable {Hamiltonian} systems connected with isospectral deformations},
  journal      = AM,
  volume       = {16},
  year         = {1975},
  pages        = {197},
  doi          = {10.1016/0001-8708(75)90151-6}
}

@article{woj83,
  author       = {Wojciechowski, S.},
  title        = {Superintegrability of the {Calogero-Moser} system},
  journal      = PLA,
  volume       = {95},
  year         = {1983},
  pages        = {279},
  doi          = {10.1016/0375-9601(83)90018-X}
}

@article{kuznetsov,
  author       = {Kuznetsov, V. B.},
  title        = {Hidden symmetry of the quantum {Calogero–Moser} system},
  journal      = PLA,
  volume       = {218},
  year         = {1996},
  pages        = {212},
  doi          = {10.1016/0375-9601(96)00412-9},
  eprint       = {solv-int/9509001},
  archivePrefix= {arXiv}
}

@article{gonera,
  author       = {Gonera, C.},
  title        = {{A note on superintegrability of the quantum Calogero model}},
  journal      = PLA,
  volume       = {237},
  year         = {1998},
  pages        = {365},
  doi          = {10.1016/S0375-9601(98)00903-7}
}

@article{rev-olsh-2,
  author       = {Olshanetsky, M. A. and Perelomov, A. M.},
  title        = {{Quantum integrable systems related to Lie algebras}},
  journal      = PRep,
  volume       = {94},
  year         = {1983},
  pages        = {313},
  doi          = {10.1016/0370-1573(83)90018-2}
}

@article{rev-poly,
  author       = {Polychronakos, A. P.},
  title        = {Physics and mathematics of {Calogero} particles},
  journal      = JPA,
  volume       = {39},
  year         = {2006},
  pages        = {12793},
  eprint       = {hep-th/0607033},
  doi          = {10.1088/0305-4470/39/41/S07},
  archivePrefix= {arXiv}
}

@article{poly92,
  author       = {Polychronakos, A.},
  title        = {Exchange operator formalism for integrable systems of particles},
  journal      = PRL,
  volume       = {69},
  year         = {1992},
  pages        = {703},
  doi          = {10.1103/PhysRevLett.69.703},
  eprint       = {hep-th/9202057},
  archivePrefix= {arXiv}
}

@article{brink,
  author       = {Brink, L. and Hansson, T. and Vasiliev, M.},
  title        = {Explicit solution to the {$N$}-body {C}alogero problem},
  journal      = PLB,
  volume       = {286},
  year         = {1992},
  pages        = {109},
  doi          = {10.1016/0370-2693(92)90166-2},
  eprint       = {hep-th/9206049},
  archivePrefix= {arXiv}
}

@article{dunkl,
  author       = {Dunkl, C. F.},
  title        = {Differential-difference operators associated to reflection groups},
  journal      = TAMS,
  volume       = {311},
  year         = {1989},
  pages        = {167},
  doi          = {10.2307/2001022}
}

@article{feigin,
  author       = {Feigin, M.},
  title        = {Intertwining relations for the spherical parts of generalized {Calogero} operators},
  journal      = {Theor. Math. Phys.},
  volume       = {135},
  year         = {2003},
  pages        = {497},
  doi          = {10.1023/A:1023231402145}
}

@article{khare,
  author       = {Khare, A.},
  title        = {Exact solution of an {$N$}-body problem in one dimension},
  journal      = JPA,
  volume       = {29},
  year         = {1996},
  pages        = {L45},
  doi          = {10.1088/0305-4470/29/19/029},
  eprint       = {hep-th/9510096},
  archivePrefix= {arXiv}
}

@article{khare99,
  author       = {Ghosh, P. K. and Khare, A.},
  title          = {{Relationship between the energy eigenstates of Calogero-Sutherland models with oscillator and Coulomb-like potentials}},
  journal      = JPA,
  volume       = {32},
  year         = {1999},
  pages        = {2129},
  doi          = {10.1088/0305-4470/32/11/008},
  eprint       = {solv-int/9808005},
  archivePrefix= {arXiv}
}

@article{flp,
  author       = {Feigin, M. and Lechtenfeld, O. and Polychronakos, A.},
  title        = {{The quantum angular Calogero--Moser model}},
  journal      = JHEP,
  year         = {2013},
  volume       = {07},
  pages        = {162},
  doi           = {10.1007/JHEP07(2013)162},
  eprint       = {1305.5841},
  archivePrefix= {arXiv},
   primaryClass = {math-ph}
}

@article{cherednik,
  author       = {Cherednik, I.},
  title        = {{A unification of Knizhnik–Zamolodchikov and Dunkl operators via affine Hecke algebras}},
  journal      = IM,
  volume       = {106},
  year         = {1991},
  pages        = {411--431},
  doi          = {10.1007/BF01243918}
}

@article{etingof,
  author       = {Etingof, P. and Ginzburg, V.},
  title        = {{Symplectic reflection algebras, Calogero--Moser space, and deformed Harish–Chandra homomorphism}},
  journal      = IM,
  volume       = {147},
  year         = {2002},
  pages        = {243},
  doi            = {10.1007/s002220100171},
  eprint       = {math/0011114},
  archivePrefix= {arXiv}
}

@book{thyssen-book,
  title={{Shattered Symmetry: Group Theory From the Eightfold Way to the Periodic Table}},
  author={Thyssen, P. and Ceulemans, A.},
  isbn={9780190611408},
  url={https://books.google.am/books?id=_lzODQAAQBAJ},
  year={2017},
  publisher={Oxford University Press}
}

@article{barut71,
  author       = {Barut, A. O. and Bornzin, G. L.},
  title        = {{SO(4,2)-formulation of the symmetry breaking in relativistic Kepler problems with or without magnetic charges}},
  journal      = JMP,
  volume       = {12},
  year         = {1971},
  pages        = {841--846},
  doi          = {10.1063/1.1665653}
}

@book{barut-book,
  title={{Theory of Group Representations and Applications}},
  author={Raczka, R. and Barut, A.O.},
  isbn={9789813103870},
  url={https://books.google.am/books?id=DAU8DQAAQBAJ},
  year={1986},
  publisher={World Scientific Publishing Company}
}

@article{fh15,
  author   = {Feigin, M. and Hakobyan, T.},
  title      = {{On Dunkl angular momenta algebra}},
  journal    = JHEP,
  volume       = {11},
  year         = {2015},
  pages        = {107},
  doi          = {10.1007/JHEP11(2015)107},
  eprint       = {1409.2480},
  archivePrefix= {arXiv},
   primaryClass = {math-ph}
}

@article{fh22,
  author       = {Feigin, M. and Hakobyan, T.},
  title        = {{Algebra of Dunkl Laplace–Runge–Lenz vector}},
  journal      = LMP,
  volume       = {112},
  year         = {2022},
  pages        = {59},
  doi            = {10.1007/s11005-022-01551-0},
  eprint       = {1907.06706},
  archivePrefix= {arXiv},
   primaryClass = {math-ph}
}

@article{runge,
  title = {{Runge-Lenz vector in the Calogero-Coulomb problem}},
  author = {Hakobyan, Tigran and Nersessian, Armen},
  journal = PRA,
  volume = {92},
  issue = {2},
  pages = {022111},
  numpages = {6},
  year = {2015},
  month = {Aug},
  publisher = {American Physical Society},
  doi = {10.1103/PhysRevA.92.022111},
  url = {https://link.aps.org/doi/10.1103/PhysRevA.92.022111},
  eprint = {1504.00760},
  archivePrefix= {arXiv},
   primaryClass = {hep-th}
}

@book{book-etin,
  author    = {Pavel Etingof},
  title     = {{Calogero--Moser Systems and Representation Theory}},
  publisher = {European Mathematical Society},
  series    = {Zurich Lectures in Advanced Mathematics},
  year      = {2005},
  isbn      = {9783037190172},
  doi       = {10.4171/034},
  url       = {https://ems.press/books/zlam/30}
}

@book{book-arut,
  author    = {Gleb Arutyunov},
  title     = {Elements of Classical and Quantum Integrable Systems},
  publisher = {Springer},
  series    = {UNITEXT for Physics},
  year      = {2019},
  isbn      = {9783030241971},
  doi       = {10.1007/978-3-030-24198-8},
  url       = {https://link.springer.com/book/10.1007/978-3-030-24198-8}
}

@article{hak14-1,
  author       = {Hakobyan, T. and Lechtenfeld, O. and Nersessian, A.},
  title        = {{Superintegrability of generalized Calogero models with oscillator or Coulomb potential}},
  journal      = PRD,
  volume       = {90},
  year         = {2014},
  pages        = {101701},
  doi          = {10.1103/PhysRevD.90.101701},
  eprint       = {1409.8288},
  archivePrefix= {arXiv},
   primaryClass = {hep-th}
}

@article{hak14-2,
    author = "Hakobyan, Tigran and Karakhanyan, David and Lechtenfeld, Olaf",
    title = {The structure of invariants in conformal mechanics},
    eprint = "1402.2288",
    archivePrefix = "arXiv",
    primaryClass = "hep-th",
    doi = "10.1016/j.nuclphysb.2014.07.008",
    journal = NPB,
    volume = "886",
    pages = "399--420",
    year = "2014"
}

@ARTICLE{hak-chain,
    author = {{Hakobyan}, Tigran},
    title = {{Symmetries of the generalized Calogero model and the Polychronakos-Frahm chain}},
    journal = PRD,
    year = 2019,
    month = may,
    volume = {99},
    number = {10},
    eid = {105011},
    pages = {105011},
    doi = {10.1103/PhysRevD.99.105011},
    archivePrefix = {arXiv},
    eprint = {1903.10030},
    primaryClass = {hep-th},
}

@article{hak23,
    doi = {10.1088/1751-8121/adaf85},
    url = {https://doi.org/10.1088/1751-8121/adaf85},
    year = {2025},
    month = {feb},
    publisher = {IOP Publishing},
    volume = {58},
    number = {6},
    pages = {065201},
    author = {Hakobyan, Tigran},
    title = {{Dunkl symplectic algebra in generalized Calogero models}},
    journal = JPA,
    eprint       = {2306.17677},
    archivePrefix = {arXiv},
    primaryClass = {hep-th}
}

@article{barut67,
  author    = {A. O. Barut and Hagen Kleinert},
  title     = {Transition Probabilities of the Hydrogen Atom from Noncompact Dynamical Groups},
  journal   = PRev,
  volume    = {156},
  pages     = {1541--1545},
  year      = {1967},
  publisher = {American Physical Society},
  doi       = {10.1103/PhysRev.156.1541},
  url       = {https://link.aps.org/doi/10.1103/PhysRev.156.1541}
}

@ARTICLE{hak16-1,
    author = {{Correa}, Francisco and {Hakobyan}, Tigran and {Lechtenfeld}, Olaf and {Nersessian}, Armen},
    title = {{Spherical Calogero model with oscillator/Coulomb potential: Classical case}},
    journal =PRD,
    year = 2016,
    month = jun,
    volume = {93},
    number = {12},
    eid = {125008},
    pages = {125008},
    doi = {10.1103/PhysRevD.93.125008},
    archivePrefix = {arXiv},
    eprint = {1604.00026},
    primaryClass = {hep-th},
}

@ARTICLE{hak16-2,
    author = {{Correa}, Francisco and {Hakobyan}, Tigran and {Lechtenfeld}, Olaf and {Nersessian}, Armen},
    title = "{Spherical Calogero model with oscillator/Coulomb potential: Quantum case}",
    journal = PRD,
    year = 2016,
    month = jun,
    volume = {93},
    number = {12},
    eid = {125009},
    pages = {125009},
    doi = {10.1103/PhysRevD.93.125009},
    archivePrefix = {arXiv},
    eprint = {1604.00027},
    primaryClass = {hep-th},
}

@ARTICLE{correa23,
    author = {{Correa}, Francisco and {Leal}, Gonzalo and {Lechtenfeld}, Olaf and {Marquette}, Ian},
    title = {{A Casimir operator for a Calogero $W$ algebra}},
    journal = JPA,
    year = 2024,
    month = feb,
    volume = {57},
    number = {8},
    eid = {085203},
    pages = {085203},
    doi = {10.1088/1751-8121/ad24ca},
    archivePrefix = {arXiv},
    eprint = {2308.07390},
    primaryClass = {hep-th},
}

@book{rev-rosler,
  author    = {M. Rösler},
  title     = {{Dunkl Operators: Theory and Applications}},
  series    = {Lecture Notes in Mathematics},
  publisher = {Springer},
  year      = {2003},
  isbn      = {978-3-540-00816-1},
  eprint   = {math/0210366},
  archivePrefix = {arXiv},
  primaryClass={math.CA}
}

@incollection{rev-chalykh,
  author    = {O. Chalykh},
  title     = {{Dunkl and Cherednik Operators}},
  booktitle = {Encyclopedia of Mathematical Physics, 2nd Edition},
  publisher = {Elsevier},
  year      = {2020},
  pages     = {129--136},
  doi        = {10.1016/B978-0-323-95703-8.00060-4},
  archivePrefix = {arXiv},
  eprint = {2409.09005},
  primaryClass = {math-ph},
}

@incollection{Heckman1997,
  author    = {Heckman, G. J.},
  title     = {Dunkl operators},
  booktitle = {S\'eminaire Bourbaki: volume 1996/97, expos\'es 820–834},
  series    = {Astérisque},
  number    = {245},
  pages     = {223--246},
  year      = {1997},
  note      = {Exposé no. 828},
  publisher = {Société Mathématique de France},
  url       = {https://www.numdam.org/item/SB_1996-1997__39__223_0/}
}

@misc{etingof-ma,
  author    = {Pavel Etingof and Xiaoguang Ma},
  title     = {{Lecture Notes on Cherednik Algebras}},
  eprint    = {1001.0432},
  year      = {2010},
  archivePrefix = {arXiv},
  primaryClass={math.RT},
  url       = {https://math.mit.edu/~etingof/18735.pdf}
}

@book{book-dunkl,
  place={Cambridge},
  edition={2},
  series={Encyclopedia of Mathematics and its Applications (155)},
  title={Orthogonal Polynomials of Several Variables},
  publisher={Cambridge University Press},
  author={Dunkl, Charles F. and Xu, Yuan},
  year={2014},
  collection={Encyclopedia of Mathematics and its Applications},
  isbn={9781107071896},
  doi={10.1017/CBO9781107786134}
}

@article{Xu-2000,
    title={Harmonic Polynomials Associated With Reflection Groups},
    volume={43},
    doi={10.4153/CMB-2000-057-2},
    number={4},
    journal=CMB,
    author={Xu, Yuan},
    year={2000},
    pages={496–507}
}

@article{isakov,
   title={{Algebra of one-particle operators for the Calogero model}},
   volume={463},
   ISSN={0550-3213},
   url={http://dx.doi.org/10.1016/0550-3213(96)00010-7},
   DOI={10.1016/0550-3213(96)00010-7},
   number={1},
   journal=NPB,
   publisher={Elsevier BV},
   author={Isakov, Serguei B. and Leinaas, Jon Magne},
   year={1996},
   month=mar,
   pages={194–214},
   eprint= {hep-th/9510184},
   archivePrefix={arXiv},
}

@article{herranz25,
    title = {{An infinite family of Dunkl type superintegrable curved Hamiltonians through coalgebra symmetry:
        Oscillator and Kepler–Coulomb models}},
    journal = PhysD,
    volume = {483},
    pages = {134963},
    year = {2025},
    issn = {0167-2789},
    doi = {https://doi.org/10.1016/j.physd.2025.134963},
    url = {https://www.sciencedirect.com/science/article/pii/S0167278925004403},
    author = {Francisco J. Herranz and Danilo Latini},
    eprint={2507.03425},
    archivePrefix={arXiv},
}

@article{serban24,
  author    = {Jules Lamers and Didina Serban},
  title     =  {{From fermionic spin-Calogero-Sutherland models to the Haldane-Shastry chain by freezing}},
  journal   = JPA,
  volume    = {57},
  year      = {2024},
  number    = {23},
  pages     = {235205},
  doi       = {10.1088/1751-8121/ad2c7a},
  eprint    = {2212.01373},
  archivePrefix = {arXiv},
  primaryClass = {math-ph},
}

@misc{chalykh24,
      title={{Integrability of the Inozemtsev spin chain}},
      author={Oleg Chalykh},
      year={2024},
      eprint={2407.03276},
      archivePrefix={arXiv},
      primaryClass={nlin.SI},
      url={https://arxiv.org/abs/2407.03276},
}

@misc{maria25,
      title={{$R$-matrix Dunkl operators and spin Calogero-Moser system}},
      author={Oleg Chalykh and Maria Matushko},
      year={2025},
      eprint={2509.18989},
      archivePrefix={arXiv},
      primaryClass={math.QA},
      url={https://arxiv.org/abs/2509.18989},
}

@ARTICLE{chalykh19,
    author = {{Chalykh}, Oleg},
    title = {{Quantum Lax pairs via Dunkl and Cherednik operators}},
    journal = CMP,
    year = 2019,
    month = jul,
    volume = {369},
    number = {1},
    pages = {261-316},
    doi = {10.1007/s00220-019-03289-8},
    archivePrefix = {arXiv},
    eprint = {\allowbreak 1804.01766},
    primaryClass = {math.QA},
}

@ARTICLE{veselov94,
    author = {{Buchstaber}, V.~M. and {Felder}, Giovanni and {Veselov}, A.~V.},
    title = {{Elliptic Dunkl operators, root systems, and functional equations}},
    journal   = {Duke Math. J.},
    volume    = {76},
    number    = {4},
    pages     = {885--911},
    year      = {1994},
    doi = {10.1215/S0012-7094-94-07635-7},
    archivePrefix = {arXiv},
    eprint = {hep-th/9403178},
    primaryClass = {hep-th},
}

@ARTICLE{cherednik95,
    author = {{Cherednik}, Ivan},
    title = {{Elliptic quantum many-body problem and double affine Knizhnik-Zamolodchikov equation}},
    journal = CMP,
    year = 1995,
    month = may,
    volume = {169},
    number = {2},
    pages = {441--461},
    doi = {10.1007/BF02099480},
    archivePrefix = {arXiv},
    eprint = {hep-th/9403136},
    primaryClass = {hep-th},
}

@ARTICLE{bernard93,
    author = {{Bernard}, D. and {Gaudin}, M. and {Haldane}, F.~D.~M. and {Pasquier}, V.},
    title = {{Yang-Baxter equation in long-range interacting systems}},
    journal =JPA,
    year = 1993,
    month = oct,
    volume = {26},
    number = {20},
    pages = {5219-5236},
    doi = {10.1088/0305-4470/26/20/010},
    archivePrefix = {arXiv},
    eprint = {hep-th/9301084},
    primaryClass = {hep-th},
}

@ARTICLE{hechman91,
    author = {{Heckman}, G.~J.},
    title = {An elementary approach to the hypergeometric shift operators of {Opdam}},
    journal = IM,
    year = 1991,
    month = dec,
    volume = {103},
    pages = {341},
    doi = {10.1007/BF01239517},
}

@ARTICLE{poly93,
       author = {{Polychronakos}, Alexios P.},
        title = {{Lattice integrable systems of Haldane-Shastry type}},
      journal = PRL,
         year = 1993,
        month = apr,
       volume = {70},
       number = {15},
        pages = {2329-2331},
          doi = {10.1103/PhysRevLett.70.2329},
archivePrefix = {arXiv},
       eprint = {hep-th/9210109},
 primaryClass = {cond-mat},
}

@ARTICLE{minahan93,
    author = {{Fowler}, Michael and {Minahan}, Joseph A.},
    title = {{Invariants of the Haldane-Shastry SU(N) chain}},
    journal = PRL,
    year = 1993,
    month = apr,
    volume = {70},
    number = {15},
    pages = {2325-2328},
    doi = {10.1103/PhysRevLett.70.2325},
    archivePrefix = {arXiv},
    eprint = {cond-mat/9208016},
    primaryClass = {cond-mat},
}

@ARTICLE{resh25,
    author = {{Liashyk}, Andrii and {Reshetikhin}, Nicolai and {Sechin}, Ivan},
    title = {Quantum Integrable Systems on a Classical Integrable Background},
    journal = CMP,
    year = 2025,
    month = dec,
    volume = {407},
    number = {1},
    eid = {15},
    pages = {15},
    doi = {10.1007/s00220-025-05523-y},
    archivePrefix = {arXiv},
    eprint = {2405.17865},
    primaryClass = {math-ph},
}

@ARTICLE{hikami,
    author = {{Ujino}, Hideaki and {Hikami}, Kazuhiro and {Wadati}, Miki},
    title = {Integrability of the Quantum {Calogero-Moser} Model},
    journal = {J. Phys. Soc. Jpn.},
    year = 1992,
    month = oct,
    volume = {61},
    number = {10},
    pages = {3425},
    doi = {10.1143/JPSJ.61.3425},
}

@ARTICLE{mathieu01,
    author = {{Mathieu}, P. and {Xudous}, Y.},
    title = {{Conserved charges of non-Yangian type for the Frahm-Polychronakos spin chain}},
    journal = JPA,
    year = 2001,
    month = may,
    volume = {34},
    number = {19},
    pages = {4197-4215},
    doi = {10.1088/0305-4470/34/19/316},
    archivePrefix = {arXiv},
    eprint = {hep-th/0008036},
    primaryClass = {hep-th},
}

@ARTICLE{talstra95,
    author = {{Talstra}, J.~C. and {Haldane}, F.~D.~M.},
    title = {{Integrals of motion of the Haldane-Shastry model}},
    journal = JPA,
    year = 1995,
    month = apr,
    volume = {28},
    number = {8},
    pages = {2369-2377},
    doi = {10.1088/0305-4470/28/8/027},
    archivePrefix = {arXiv},
    eprint = {cond-mat/9411065},
    primaryClass = {cond-mat},
}

@ARTICLE{hak-kriv,
       author = {{Hakobyan}, Tigran and {Krivonos}, Sergey and {Lechtenfeld}, Olaf and {Nersessian}, Armen},
        title = {Hidden symmetries of integrable conformal mechanical systems},
      journal = PLA,
         year = 2010,
        month = jan,
       volume = {374},
       number = {6},
        pages = {801-806},
          doi = {10.1016/j.physleta.2009.12.006},
archivePrefix = {arXiv},
       eprint = {0908.3290},
 primaryClass = {hep-th},
}

@article{fubini,
    author = "de Alfaro, Vittorio and Fubini, S. and Furlan, G.",
    title = {Conformal Invariance in Quantum Mechanics},
    reportNumber = "CERN-TH-2115",
    doi = "10.1007/BF02785666",
    journal = "Nuovo Cim. A",
    volume = "34",
    pages = "569",
    year = "1976"
}

@misc{feigin23,
      title={{$q$-Analogue of the degree zero part of a rational Cherednik algebra}},
      author={Misha Feigin and Martin Vrabec},
      year={2024},
      eprint={2311.07543},
      archivePrefix={arXiv},
      primaryClass={math.QA},
      url={https://arxiv.org/abs/2311.07543},
}

@phdthesis{vrabec25,
  author       = {Vrabec, Martin},
  title        = {Quantum many-body integrable systems and related algebraic structures},
  school       = {University of Glasgow},
  year         = {2025},
  type         = {{PhD thesis}},
  doi           =  {10.5525/gla.thesis.85399},
  url          = {https://theses.gla.ac.uk/85399/}
}

@article{feigin26,
    title = {{Two invariant subalgebras of rational Cherednik algebras}},
    journal = JPAA,
    volume = {230},
    number = {3},
    pages = {108190},
    year = {2026},
    issn = {0022-4049},
    doi = {https://doi.org/10.1016/j.jpaa.2026.108190},
    url = {https://www.sciencedirect.com/science/article/pii/S0022404926000216},
    author = {Gwyn Bellamy and Misha Feigin and Niall Hird},
    archivePrefix={arXiv},
    primaryClass={math.QA},
    eprint={2312.13957},
}

@article{turbiner,
    author = "Turbiner, Alexander",
    title = {{Hidden algebra of the $N$ body Calogero problem}},
    eprint = "hep-th/9310125",
    archivePrefix = "arXiv",
    doi = "10.1016/0370-2693(94)90657-2",
    journal = PLB,
    volume = "320",
    pages = "281--286",
    year = "1994"
}

@ARTICLE{mikhail,
    author = {{Correa}, Francisco and {Lechtenfeld}, Olaf and {Plyushchay}, Mikhail},
    title = {{Nonlinear supersymmetry in the quantum Calogero model}},
    journal = JHEP,
    year = 2014,
    month = apr,
    volume = {2014},
    eid = {151},
    pages = {151},
    doi = {10.1007/JHEP04(2014)151},
    archivePrefix = {arXiv},
    eprint = {1312.5749},
    primaryClass = {hep-th},
}

\appendix

\section{Commutation relations in the $\deform{so(2,N)}$ algebra}
\label{app:comm}
This appendix is devoted to the derivation of certain commutation relations
among the $\deform{so(N+1,2)}$ algebra generators introduced in Sect.~\ref{sec:dyn}.

First, note that the square of the Dunkl momentum satisfies the following commutation relations:
\begin{gather}
[\pi^2,f(r)]=[p_r^2,f(r)]=-f''(r)-2\imath f'(r)p_r,
\qquad
[\pi^2,x_i]=-2\imath\pi_i.
\end{gather}
Using these relations, together with the definitions of the conformal generators
\eqref{K123}, it is straightforward to verify that the following commutator rules
hold for $\sigma=1,2$:
\begin{align}
\label{comKsr}
[K_\sigma,r] &=-\imath rp_r =-\imath K_3,
&
[K_\sigma,x_i]&=-\imath r\pi_i,
\\
\label{comKsxr}
\Big[K_\sigma,\frac{x_i}r\Big]
&=-\imath\pi_i+\frac{x_i}{r^2}\imath K_3,
&
[K_\sigma,\pi_i]&=\imath\frac{x_i}{r^2} K_\sigma,
\end{align}

We are now ready to calculate the commutation relations of the elements $K_\sigma$
with the components of the three (Dunkl) vectors $A_\sigma$ and $\Gamma$ appearing in the
matrix $\gen$ \eqref{genMat}.
Indeed, Eqs.~\eqref{comKsAG} hold, since we have ($\sigma, \sigma_1, \sigma_2 = 1,2$):
\begin{align}
\label{comKsA}
\nonumber
[K_{\sigma_1},A_{\sigma_2}]
      &=\left[K_{\sigma_1},\frac{x}{r}\right] K_{\bar\sigma_2} + \frac{x}{r}[K_{\sigma_1},K_{\bar\sigma_2}]
        -[K_{\sigma_1},K_3]\pi - K_3
\\
\nonumber
      &=\left(-\imath\pi+\frac{x}{r^2}\imath K_3\right)K_{\bar\sigma_2} +  \epsilon_{\sigma_1\bar\sigma_2}\frac{x}{r} \imath K_3
         +K_{\bar\sigma_1}\imath\pi- \imath K_3 \frac{x}{r^2} K_{\sigma_1}
\nonumber
\\
      &=(K_{\bar\sigma_1}-K_{\bar\sigma_2})\imath\pi + \frac{x}{r^2} ( K_{\sigma_1} - K_{\bar\sigma_2})
         + \frac{x}{r^2}\imath K_3 (K_{\bar\sigma_2}-K_{\sigma_1})
         + \epsilon_{\sigma_1\bar\sigma_2}\frac{x}{r} \imath K_3
 \nonumber
\\
      &=\imath\epsilon_{\bar\sigma_1\bar\sigma_2} r\pi + \epsilon_{\sigma_1\bar\sigma_2}\frac{x}{r}
         + \epsilon_{\bar\sigma_2\sigma_1} \frac{x}{r^2}\imath K_3  r
         + \epsilon_{\sigma_1\bar\sigma_2}\frac{x}{r} \imath K_3
 =-\imath\epsilon_{\sigma_1\sigma_2}\Gamma,
\\[3pt]
\label{comKsG}
[K_\sigma,\Gamma]
            &= [K_{\sigma},r] \pi + r[K_\sigma,\pi]
            =-\imath K_3\pi+  \imath\frac{x}{r} K_\sigma
            =\imath A_{\bar\sigma } .
\end{align}
In the above derivation, Eqs.~\eqref{comKsr} and \eqref{comKsxr}, together
with the definitions \eqref{K123} and \eqref{Krad}, have been used.
Note that the coordinate index $i$ for all vectors is omitted for clarity.

We now turn to the commutation relations among the components of the
vectors $A_\alpha$. For the third vector ($\alpha=3$), the commutators are straightforward, since:
\begin{equation}
\label{comGG'}
\begin{aligned}
{} [\Gamma_i,\Gamma_j]&=[r\pi_i,r\pi_j]=r[\pi_i,r]\pi_j - r [\pi_j,r]\pi_j=-\imath (x_i\pi_j - x_j\pi_i)
=-\imath L_{ij},
\end{aligned}
\end{equation}
which proves Eq.~\eqref{comGG}.
 Next, we focus on the
mixed commutator \eqref{comAsG}.
Using   the representation \eqref{AsigK} of the vector $A_\sigma$, we get:
\begin{equation}
\label{comAsG'}
\begin{aligned}
{} [A_{\sigma i},\Gamma_j]&=
\left[\frac{x_i}r,\Gamma_j\right]K_{\bar\sigma}
    +\frac{x_i}r[K_{\bar\sigma},\Gamma_j]-K_3[\pi_i,\Gamma_j]
    =\left(\imath S_{ij}-\frac{\imath x_ix_j}{r^2}\right)K_{\bar \sigma} + \frac{\imath x_i}r \left(A_{\sigma j}+K_3\pi_j\right)
    = \imath K_{\bar\sigma}S_{ij}.
\end{aligned}
\end{equation}
The second equation above employs the previously derived relation \eqref{comKsG}.

Finally, consider the commutation relations between the components
of  $A_\sigma$. For this calculation, it is convenient to use the following representation:
\begin{equation}
A_{\sigma i}=x_ib_{\bar\sigma}+K_3\pi_i,
\qquad
\text{where}
\qquad
b_\sigma=\frac1rK_\sigma=\frac12\left(\pi^2+(-1)^{\bar \sigma}\right).
\end{equation}
Then, the operators $\pi_k$ and $b_\sigma$ mutually commute,
and the commutators between the different terms can be readily computed:
\begin{align}
\label{comAA-22}
[K_3\pi_i,K_3\pi_j]&=K_3[\pi_i,K_3]\pi_j+K_3[K_3,\pi_j]\pi_i=-\imath K_3\pi_i\pi_j+\imath K_3\pi_j\pi_i=0,
\\[2pt]
\label{comAA-11}
[x_i b_{\bar\sigma_1},x_jb_{\bar\sigma_2}]&=
x_i [b_{\bar\sigma_1},x_j] b_{\bar\sigma_2}  -  x_j [b_{\bar\sigma_2},x_i] b_{\bar\sigma_1}
=-\imath(x_i\pi_j b_{\bar\sigma_2} -x_j\pi_i b_{\bar\sigma_1}),
\\[2pt]
\label{comAA-12}
[K_3\pi_j,x_ib_{\sigma}]&=[K_3,x_i]b_{\sigma}\pi_j +x_i  [K_3,b_{\sigma}]\pi_j
+K_3[\pi_j,x_i]b_{\sigma}
\nonumber
\\
&
=\imath x_i(\pi^2-b_{\sigma})\pi_j-\imath K_3 b_{\sigma} S_{ij}
=\imath x_i\pi_j b_{\bar \sigma}-\imath K_3 b_{\sigma} S_{ij}.
\end{align}
As a consequence of the above relations, we obtain:
\begin{align}
\label{comAA'}
\big[A_{\sigma_1 i},A_{\sigma_2 j}\big]&=[x_i b_{\bar\sigma_1},x_j b_{\bar\sigma_2}]
+[K_3\pi_i,x_j b_{\bar\sigma_2}]-[K_3\pi_j,x_i b_{\bar\sigma_1}]
\nonumber\\
&=
-\imath(x_i\pi_j b_{\bar\sigma_2} -x_j\pi_i b_{\bar\sigma_1})
+\imath (x_i\pi_j b_{\sigma_1} - x_j\pi_i b_{\sigma_2} )
-\imath K_3 (b_{\bar\sigma_1}- b_{\bar\sigma_2}) S_{ij}
\nonumber\\
&=\imath\epsilon_{\sigma_1\bar\sigma_2} L_{ij}+\imath\epsilon_{\sigma_1\sigma_2}K_3 S_{ij},
\end{align}
which proves Eq.~\eqref{comAsAs}.

\section{Squares of the three Dunkl vectors $A_\alpha$  in $\deform{so(2,N)}$}
\label{app:Asq}

In this Appendix, we derive the formulas \eqref{Asq}--\eqref{Gsq} for the squares of the three
Dunkl vectors and their combined expression \eqref{AalSq}.

Before proceeding, we recall the following simple relation \cite{fh15}, which can be verified using
Eqs.~\eqref{Sij} and Eq.~\eqref{K123}:
\begin{equation}
\label{x.pi}
x\cdot\imath\pi=r\partial_r+S=\imath K_3+S-\sfrac{N-1}2.
\end{equation}

According to the representation \eqref{AsigK} for the first two vectors ($\sigma=1,2$),
we obtain:
\begin{align}
\label{AMsq}
A_\sigma^2&= \sum_{i=1}^N \left[\left(\frac{x_i}{r}K_{\bar\sigma}\right)^2+(K_3\pi_i)^2
-\left\{\frac{x_i}{r}K_{\bar\sigma}, K_3\pi_i\right\}\right],
\end{align}
where the curly brackets denote the anticommutator.
The first two terms in the above sum  can be simplified using Eqs.~\eqref{comKsxr} and \eqref{x.pi}:
\begin{align}
\sum_{i=1}^N \left(\frac{x_i}{r}K_{\bar\sigma}\right)^2
&=K_{\bar\sigma}^2+\frac{1}{r}\left(- x\cdot \imath \pi+\imath K_3\right)K_{\bar\sigma}
=K_{\bar\sigma}^2+\frac1r\left(\sfrac{N-1}{2}-S\right)K_{\bar\sigma},
\label{AMsq1}
\\
\sum_{i=1}^N ( K_3\pi_i)^2&=-\imath K_3(\imath K_3+1)\pi^2.
\label{AMsq2}
\end{align}
The last term in \eqref{AMsq}  can be evaluated using the definitions \eqref{K123}, \eqref{Krad}, and the
commutation relations \eqref{comKs}:
\begin{equation}
\label{AMsq3}
\begin{aligned}
-\sum_{i=1}^N \big(\frac{x_i}{r}K_{\bar\sigma} K_3\pi_i &+  K_3\pi_i \frac{x_i}{r}K_{\bar\sigma} \big)
=\frac1r ( x\cdot \imath\pi-1)K_{\bar\sigma}(\imath K_3-1) + \imath K_3( x\cdot \imath \pi+N-2S) \frac1{r}K_{\bar\sigma}
\\
=&
\frac{1}r(\imath K_3+S-\sfrac{N+1}{2})\big(\imath K_3K_{\bar\sigma}+(-1)^{\bar\sigma} r\big)
 +\frac{1}r(\imath K_3-1)(\imath K_3-S+\sfrac{N-1}{2})K_{\bar\sigma}
\\
=&
\frac{1}r\big[2\imath K_3(\imath K_3-1)+S-\sfrac{N-1}{2}\big]K_{\bar\sigma}
+ (-1)^{\bar\sigma}(\imath K_3+S-\sfrac{N-1}{2}).
\end{aligned}
\end{equation}
In Eq.~\eqref{AMsq3}, the coordinate and Dunkl momentum operator are reordered
 to form the scalar product, which is expressed in terms of $K_3$  via \eqref{x.pi}.
Summing the contributions \eqref{AMsq1}--\eqref{AMsq3},  we obtain:
\begin{equation}
\label{Asq'}
\begin{aligned}
A_\sigma^2&=K_{\bar\sigma}^2+ \frac2r \imath K_3(\imath K_3-1)K_{\bar\sigma}-\imath K_3(\imath K_3+1)\pi^2+(-1)^{\bar\sigma}(\imath K_3+S-\sfrac{N-1}{2})
\\
&=K_{\bar\sigma}^2-\imath K_3\big(\imath K_3+1)(\pi^2+(-1)^{\bar\sigma}\big)-\imath K_3(\imath K_3+1)\pi^2+(-1)^{\bar\sigma}(\imath K_3+S-\sfrac{N-1}{2})
\\
&=K_{\bar\sigma}^2+(-1)^{\bar\sigma}\big(K_3^2+S-\sfrac{N-1}{2}\big),
\end{aligned}
\end{equation}
which yields Eqs.~\eqref{Asq} and \eqref{Msq} for $\sigma=1$ and $\sigma=2$, respectively.

Finally, the expression \eqref{Gsq} for the square of the third vector \eqref{Gi} follows
from comparing the following two relations:
\begin{align}
\label{Gsq'}
\Gamma^2
&=r^2\pi^2-\imath x\cdot \pi
=r^2\pi^2-\imath K_3 -S +\sfrac{N-1}{2},
\\
K_1^2-K_2^2&=(K_1-K_2)(K_1+K_2)-[K_1,K_2]=r^2\pi^2-\imath K_3.
\end{align}

\end{document}